\documentclass[preprint,aps,showpacs,pra,superscriptaddress]{revtex4-1}
\usepackage{graphicx}
\usepackage{amsfonts}
\usepackage{amssymb}
\usepackage{amsmath}
\usepackage{color}

\begin{document}

\title{
Dark solitons in cigar-shaped Bose-Einstein condensates in double-well potentials}
\author{S.\ Middelkamp}
\affiliation{Zentrum f\"ur Optische Quantentechnologien, Universit\"at
Hamburg, Luruper Chaussee 149, 22761 Hamburg, Germany}
\author{G.\ Theocharis}
\affiliation{Department of Mathematics and Statistics, University of Massachusetts,
Amherst MA 01003-4515, USA}
\author{P.G.\ Kevrekidis}
\affiliation{Department of Mathematics and Statistics, University of Massachusetts,
Amherst MA 01003-4515, USA}
\author{D.J.\ Frantzeskakis}
\affiliation{Department of Physics, University of Athens, Panepistimiopolis, Zografos, Athens 157 84, Greece}
\author{P.\ Schmelcher}
\affiliation{Zentrum f\"ur Optische Quantentechnologien, Universit\"at
Hamburg, Luruper Chaussee 149, 22761 Hamburg, Germany}

\begin{abstract}

We study the statics and dynamics of dark solitons in a cigar-shaped Bose-Einstein condensate confined in a double-well potential.
Using a mean-field model with a non-cubic nonlinearity, appropriate to describe the dimensionality crossover regime from one to three dimensional, we obtain branches of solutions in the form of single- and multiple-dark soliton states, and study their bifurcations and stability. It is demonstrated that there 
exist dark soliton states which do not have a linear counterpart and we highlight the role of anomalous modes in the excitation spectra. 
Particularly, we show that anomalous mode eigenfrequencies are closely connected to the characteristic soliton frequencies as found 
from the solitons' equations of motion, and how anomalous modes are related to the emergence of instabilities. We also analyze in detail the 
role of the height of the barrier in the double well setting, which may lead to instabilities or decouple multiple dark soliton states.

\end{abstract}

\maketitle

\section{Introduction}

For many purposes atomic Bose-Einstein condensates (BECs) can be accurately described by 
means of a mean-field theoretical model, the Gross-Pitaevskii (GP) equation \cite{BECBook,book2,revnonlin}. Importantly the GP equation describes
not only the ground-state, but also macroscopic nonlinear excitations of BECs, such as matter-wave dark solitons, which have been studied theoretically  
(see, e.g., Refs.~\cite{revnonlin,ourbook} and references therein) and were observed in a series of experiments 
\cite{han1,nist,dutton,bpa,ginsberg2005,engels,hamburg,hambcol,kip,technion,draft6}. In fact, these localized nonlinear 
structures have attracted much attention as they arise spontaneously upon crossing the BEC phase-transition \cite{zurek2,zurek3}, 
while their properties may be used as diagnostic tools for probing properties of BECs \cite{anglin}. 
Additionally, applications of matter-wave dark solitons have been proposed: the dark soliton 
position can be used to monitor the phase acquired in an atomic matter-wave interferometer in the nonlinear regime \cite{appl1,appl2}.

As matter-wave dark solitons are known to be more robust in the quasi one-dimensional (1D) geometry, the majority of relevant theoretical 
studies have been performed in the framework of the 1D GP equation and, particularly, in the so-called 
Thomas-Fermi (TF)-1D regime (see, e.g., Ref.~\cite{kip}). More specifically, many works have been devoted to the stability 
\cite{fms,mprizolas,muryshev} and dynamical properties of dark solitons, such as their oscillations 
\cite{fms,mprizolas,muryshev,motion1,motion2,motion3,motion4,motion5,motion6,motion7} and sound emission \cite{motion5,nppsound} in the presence 
of a harmonic trapping potential. In the TF-1D regime, matter-wave dark solitons have also been studied in periodic (optical lattice) 
potentials \cite{bbbb1,yulin,bbbb2,bbbb3,dsolyuri}, as well as in combinations of harmonic traps and optical lattices \cite{weol,nppOL,gt2,Theocharis}. 
Notice that upon properly choosing the harmonic trap and optical lattice parameters, it is possible to create a {\it double-well potential}, 
as was demonstrated experimentally in Ref.~\cite{albiez}. It may also be formed by combining a harmonic trap with a repulsive 
barrier potential, induced by a blue-detuned laser beam \cite{interf}. Double-well potentials have been studied in theory  
\cite{smerzi,kiv2,mahmud,bam,Bergeman_2mode,infeld,todd,Theo06}, while they have played a key role in important experimental observations: 
these include Josephson oscillations and tunneling (for a small number of atoms) or macroscopic quantum self-trapping and an asymmetric partition 
of the atoms between the wells (for sufficiently large numbers of atoms) \cite{albiez}, as well as nonlinear matter-wave interference 
leading to the formation of matter-wave vortices \cite{bpa2,Nate2} and dark solitons \cite{kip,draft6}.

So far, there exist only a few studies on matter-wave dark solitons in double-well potentials \cite{yuridw,ichihara}. In particular, 
Ref.~\cite{yuridw} was devoted to the study of nonlinearity-assisted quantum tunneling and formation of 
dark solitons in a matter-wave interferometer, which is realized by the adiabatic transformation of a double-well potential into a 
single-well harmonic trap. On the other hand, in Ref.~\cite{ichihara}, the stability of the first excited state of a quasi-1D BEC 
trapped in a double-well potential was studied and regimes of (in)stability were found; note that in the nonlinear regime, the first excited state 
of the BEC is nothing but the stationary ``black'' soliton --- alias ``kink'' --- solution of the pertinent 1D GP equation. 
At this point, it is relevant to mention that both Refs.~\cite{yuridw,ichihara} follow the majority of the theoretical works on 
dark solitons in BECs, as they have also been performed in the TF-1D regime. Nevertheless, this regime has not been 
practically accessible in real experiments: in fact, in most relevant experiments the condensates were usually elongated 
(alias ``cigar-shaped'') three-dimensional (3D) objects, while only two recent experiments \cite{kip,draft6} were conducted 
in the so-called dimensionality crossover regime from 1D to 3D \cite{str}. The observations of these experiments were found to be in 
very good agreement with the theoretical predictions \cite{kip,draft6} (see also Ref.~\cite{Theo2007}, based on the use of 
effective 1D mean-field models devised in Refs.~\cite{gerbier,npse,Delgado}). 

In the present 
work we study dark soliton states in cigar-shaped BECs --- being in the 
dimensionality crossover regime from 3D to 1D --- confined in double-well potentials that are formed by a combination of a harmonic 
trap and an optical lattice. Our analysis relies on the use of an effective 1D mean-field model, 
which was first introduced in Ref.~\cite{gerbier} and later was rederived and analyzed in detail in Refs.~\cite{Delgado}. This model, 
which has the form of a GP-like equation with a non-cubic nonlinearity, was also successfully used in Ref.~\cite{draft6} to analyze 
dark soliton statics and dynamics observed in the experiment. Our aim is to systematically study, apart from the single stationary dark soliton 
(see Ref.~\cite{ichihara} for a study in the TF-1D regime), multiple dark soliton states as well; such a study is particularly relevant, as 
many dark solitons can be experimentally created by means of the matter-wave interference method, as demonstrated in Refs.~\cite{kip,draft6}. 
Our study starts from the non-interacting (linear) limit, where we obtain the pertinent excited states of the BEC and, then, using continuation 
in the chemical potential (number of atoms), we find all branches of purely nonlinear solutions, including the ground state and 
single or multiple dark solitons, and their bifurcations. An important finding of our analysis is that the optical lattice (which sets the 
barrier in the double-well setting) results in the emergence of nonlinear states that do not have a linear counterpart, such as 
dark soliton states, with solitons located in one well of the double-well potential. As concerns the bifurcations of the various branches of solutions, 
we show that particular states emerge or disappear for certain chemical potential thresholds, which are determined analytically, 
in some cases, by using a Galerkin-type approach \cite{Theo06}. For each branch, we also study the stability of the pertinent solutions 
via a Bogoliubov-de Gennes (BdG) analysis, which reveals the role of the anomalous 
(negative energy) modes in the excitation spectra as concerns the emergence of oscillatory instabilities of dark solitons: it is found that such instabilities 
occur due to the collision of an anomalous mode with a positive energy mode. Furthermore, based on the weakly-interacting limit of the model, 
we study analytically small-amplitude oscillations of the one- and multiple-dark-soliton states. Particularly, we obtain equations of motion 
for the single and multiple solitons from which we determine the characteristic soliton frequencies; the latter, are found to be, in many cases 
in good agreement with the eigenfrequencies found in the framework of the BdG analysis. 

The paper is organized as follows. We present our model in section II and provide the theoretical framework of our analysis. 
In Sec. III, we study the bifurcations of the branches of the solutions of the model and study their stability via the BdG equations. 
In Sec. IV, we provide some analytical estimates for the soliton frequencies based on the dynamics of solitons and compare our findings to 
the ones obtained by the BdG analysis. Sec. V is devoted to a study of the soliton dynamics numerically, and shows the manifestation of 
instabilities when they arise. Finally, in Sec.VI, we present our conclusions.

\section{Model and theoretical Framework}
\label{sec2}

\subsection{The effective 1D mean-field model and BdG analysis}

We consider a BEC confined in a highly elongated trap, with longitudinal and transverse confining frequencies (denoted by $\omega_z$ and 
$\omega_{\perp}$, respectively) such that $\omega_z \ll \omega_{\perp}$. In such a case, use of the adiabatic approximation, together with 
a variational approach for determining the local transverse chemical potential, leads to the following 1D GP equation for the longitudinal 
condensate's wave function \cite{gerbier,Delgado}:
\begin {equation}
i\hbar \partial_t \psi=
\left[-\frac{\hbar^{2}}{2m} \partial_z^{2}+ V_{\rm ext}(z)
+\hbar \omega_\perp\sqrt{1+4 a |\psi|^2} \right] \psi,
\label{gerbier}
\end {equation}
where $\psi(z,t)$ is normalized to the number of atoms, i.e., $N=\int_{-\infty}^{+\infty} |\psi|^2 dx$, $a$ is the $s$-wave scattering length, 
$m$ is the atomic mass, and $V_{ext}(z)$ is the longitudinal part of the external trapping potential. As demonstrated in Refs.~\cite{Delgado},
Eq.~(\ref{gerbier}) provides accurate results in the dimensionality crossover regime and the TF limit, thus describing the axial dynamics of 
cigar-shaped BECs in a very good approximation to the 3D GP equation. It is worth mentioning that in the weakly-interacting limit, $4a |\psi|^2 \ll 1$, 
Eq.~(\ref{gerbier}) is reduced to the usual 1D GP equation with a cubic nonlinearity, characterized by an effective 1D coupling constant 
$g_{1D} = 2 a \hbar \omega_{\perp}$ (see, e.g., discussion and references in Ref.~\cite{revnonlin}).

Let us now assume that the trapping potential is a superposition of a harmonic trap and an optical lattice, characterized by an 
amplitude $V_0$ and a wavenumber $k$, namely,
\begin{equation}
V_{\rm ext}(z) = \frac{1}{2}m\omega_z^2 z^2+V_0\cos^2(kz).
\end{equation}
It is clear that a proper choice of the potential parameters (and the number of atoms) may readily lead to a trapping potential that has the 
form of an effective double well potential; in such a setting, the height of the barrier can be tuned by the amplitude of the optical lattice. 
In our considerations below, we consider the case of a $^{87}$Rb condensate, confined in a harmonic trap with frequencies 
$\omega_\perp= 10\omega_z=2\pi \times 400$Hz (these values are relevant to the experiments of Refs.~\cite{kip,draft6}); furthermore, 
we will vary the optical lattice strength in the range of $V_0=0 $ to $V_0=1.16 \times 10^{-12}$~eV, thereby changing 
the trapping potential from a purely harmonic form to a double-well potential. Finally, we will assume a fixed value of the optical 
lattice wavenumber, namely $k=\pi/5.37$ $\mu\text{m}^{-1}$.


Next, assuming that the density, length, time and energy are measured, respectively, in units of $a$, 
$\alpha_{\perp} \equiv \sqrt{\hbar/m\omega_{\perp}}$ (transverse harmonic oscillator length), 
$\omega_{\perp}^{-1}$, and $\hbar \omega_{\perp}$, we express Eq.~(\ref{gerbier}) in the following dimensionless form,
%
%
\begin{equation}
i \partial_t \psi =\hat{H}\psi+ \sqrt{1+4|\psi|^2}\psi,
\label{1dDelgado}
\end{equation}
where $\hat{H}=-(1/2)\partial_{z}^{2} + (1/2)\Omega^2 z^2 + V_0\cos^2(kz)$, is the usual single-particle Hamiltonian and 
$\Omega \equiv \omega_z/\omega_{\perp}$ is the normalized harmonic trap strength (note that the normalized optical lattice strength $V_0$ 
is measured in the units of energy $\hbar \omega_{\perp}$).
%
%

Below, we will analyze the stability of the nonlinear modes of Eq.~(\ref{1dDelgado}) by means of the Bogoliubov-de Gennes (BdG) equations 
(see, e.g., Ref.~\cite{BECBook}). In particular, once a numerically exact --- up to a prescribed tolerance --- stationary state, $\psi_{0}(z)$, 
is found (e.g., by a Newton-Raphson method), we consider small perturbations of this state of the form,
\begin{equation}
\psi(z,t)=\left[\psi_{0}(z)+ \left(u(z)e^{-i\omega t}+\upsilon^{\ast}(z)e^{i\omega^{\ast} t}\right)\right]e^{-i\mu t},
\label{ansatz}
\end{equation}
where $\ast$ denotes complex conjugate. This way, we derive from Eq.~(\ref{1dDelgado})
the following BdG equations:
\begin{eqnarray}
&&[\hat{H} - \mu + f] u + g\upsilon = \omega u,
\label{BdG1} \\
&&[\hat{H} - \mu + f] \upsilon + gu = -\omega \upsilon,
\label{BdG2}
\end{eqnarray}
where
$\mu$ is the chemical potential, and the functions $f$ and $g$ are given by $f= g +\sqrt{1+4n_{0}}$, and 
$g = 2 \psi_{0}^2/ \sqrt{1+4n_{0}}$ (with $n_{0} \equiv |\psi_{0}|^2$). Then, solving
Eqs.~(\ref{BdG1})-(\ref{BdG2}), we determine the eigenfrequencies
$\omega \equiv \omega_{r}+i \omega_{i}$ and the amplitudes $u$ and
$\upsilon$ of the normal modes of the system. Note that if $\omega$ is an eigenfrequency
of the Bogoliubov spectrum, so are $-\omega$, $\omega^{\ast}$ and
$-\omega^{\ast}$ (due to the Hamiltonian nature of the system), and 
consequently
the occurrence of a complex eigenfrequency always leads to a dynamic instability. 
Thus, a linearly stable configuration is
tantamount to $\omega_i =0$, i.e., all eigenfrequencies being real.
An important quantity resulting from the BdG analysis is the amount of energy carried by the normal mode with eigenfrequency $\omega$, namely
%
\begin{equation}
E=\int{dz(|u|^2-|\upsilon|^2)}
\omega.
\label{energy}
\end{equation}
The sign of this quantity, known as {\it Krein sign} \cite{MacKay},
is a topological property of each eigenmode. 
Importantly, if the normal mode eigenfrequencies with 
opposite energy (Krein) signs become resonant then, in most cases, 
there appear complex frequencies in the excitation spectrum, i.e., a dynamical instability occurs \cite{MacKay}. In order to further elaborate on 
such a possibility, it is relevant to note that modes with complex or imaginary frequencies carry zero energy (due to the condition $(\omega-\omega^{\ast})\int dz (|u|^2-|\upsilon|^2)=0$, which must hold for each BdG mode), while {\it anomalous modes} have 
negative energy (see, e.g., Sec.~5.6 of Ref.~\cite{BECBook}). The presence of anomalous modes in the excitation spectrum is a direct signature of 
an energetic instability, which is particularly relevant to the case of dark solitons \cite{fms}; such excited states of the system are discussed below.
%

\subsection{Dark soliton states}

Exact analytical dark soliton solutions of NLS equations with a generalized defocusing nonlinearity, such as Eq.~(\ref{1dDelgado}), are not available. 
In fact, in such cases, dark solitons may only be found in an implicit form (via a phase-plane analysis) or in an approximate form (via the 
small-amplitude approximation) \cite{bass}. Nevertheless, exact analytical dark soliton solutions of Eq.~(\ref{1dDelgado}) can be found in the 
weakly-interacting limit ($4|\psi|^2 \ll 1$), and in the absence of the external potential: in this case, Eq.~(\ref{1dDelgado}) is reduced to the 
completely integrable cubic NLS model, which possesses single- and multiple-dark-soliton solutions on top of a background with constant density 
$n=n_0=\mu$ (with $\mu$ being the chemical potential). A single dark soliton solution has the form \cite{zsd},
\begin{equation}
\psi(z,t) = \sqrt{n_0}\left[ i\nu +B\tanh(\eta)  \right] \exp(-i\mu t),
\label{single}
\end{equation}
where $\eta = \sqrt{n_0}B[z-z_0(t)]$, $z_0(t)$ is the soliton center, the parameter $B\equiv \sqrt{1-\nu^2}$ sets the soliton depth given 
by $\sqrt{n_0} B$, while the parameter $\nu$ sets the soliton velocity, given by $dz_0/dt=\sqrt{n_0}\nu$. Note that for $\nu=0$ the dark 
soliton becomes a black soliton (alias a stationary kink), with a $\pi$ phase jump across its density minimum.
%
Aside from the single-dark-soliton, multiple dark soliton solutions of the cubic NLS equation are also available \cite{zsd,Blow,AA}. 
In the simplest case of a two-soliton solution, with the two solitons moving with opposite velocities, $\nu_1=-\nu_2=\nu$, the wave function can 
be expressed as \cite{AA} (see also Ref.~\cite{gagnon}):
\begin{equation}
\psi(z,t) = \frac{F(z,t)}{G(z,t)}\exp(-i\mu t),
\label{double}
\end{equation}
where $F=2(n_0-2n_{\rm min})\cosh(T)-2n_0 \nu \cosh(Z)+i\sinh(T)$, $G=2\sqrt{n_0}\cosh(T)+2 \sqrt{n_{\rm min}} \cosh(Z)$, 
while $Z=2\sqrt{n_0} B z$, $T=2\sqrt{n_{\rm min}(n_0 - n_{\rm min})}t$, and $n_{\rm min} =n_0 -n_0 B^2= n_0 \nu^2$ is the minimum density 
(i.e., the density at the center of each soliton).

Generally, the single dark soliton, as well as all higher-order dark soliton states, can be obtained in a stationary form from the 
{\it non-interacting} (linear) limit of Eq.~(\ref{1dDelgado}), corresponding to $N \rightarrow 0$ \cite{KivsharPLA,konotop1}. In this limit, 
Eq.~(\ref{1dDelgado}) is reduced to a linear Schr{\"o}dinger equation for a confined single-particle state, namely
the equation of the quantum harmonic oscillator, together with the contribution
of the optical lattice. However, due to the presence of the harmonic trap, the problem
is characterized by discrete energy levels and corresponding localized eigenmodes. In the
{\it weakly-interacting} limit, where Eq.~(\ref{1dDelgado}) is reduced to the cubic NLS equation, all these eigenmodes exist for 
the nonlinear problem as well \cite{KivsharPLA,konotop1}, describing an analytical continuation of the linear modes to a set of nonlinear 
stationary states. Recent analysis and numerical results \cite{AZ} (see also Ref.~\cite{draft6}) suggest that there are no solutions of 
Eq.~(\ref{1dDelgado}) without a linear counterpart, at least in the case of a purely harmonic potential.
However, as we will show below, the presence of the optical lattice in the model results in the occurrence of 
additional states, for sufficiently large nonlinearity (large atom numbers), which do not have a linear counterpart. 

Next, let us discuss the dynamics of dark solitons in the considered setup with an external potential in the weakly-interacting limit ($4|\psi|^2 \ll 1$). Therefore, we factorize the total wavefunction $\psi=n_0 u$ into the background density $n_0$ and the soliton wavefunction $u$. We assume that the background density is given by the ground state wavefunction of the condensate and that we can use the TF approximation to describe the latter. Then  Eq.~(\ref{1dDelgado}) may be expressed as a perturbed NLS equation for the soliton wavefunction $u$, namely,
\begin{equation}
i \partial_t u + \frac{1}{2}\partial_z^2 u -\mu (|u|^2-1)u = P(u),
\end{equation}
where the effective perturbation $P(u)$ is given by:
\begin{equation}
P(u)=(1-|u|^2)Vu+ \frac{1}{2}\frac{1}{\mu - V(z)}\frac{dV}{dz}\frac{\partial u}{\partial z}.
\end{equation}
Here, $V(z)=(1/2)\Omega^2 z^2+ V_0\cos^2(kz)$ is the trapping potential, and all terms of $P(u)$
are assumed to be of the same order.
Furthermore, assuming that the length scale of the trap is much larger than the width of the soliton,
one can apply the adiabatic perturbation theory for dark solitons devised in Ref.~\cite{KivsharPRE,motion2} to derive 
the following equation of motion for the slowly-varying dark soliton center $z_0(t)$
\cite{Theocharis}:
\begin{equation}
\frac{d^2 z_0}{dt^2} = -\frac{1}{2}\frac{dV_{\rm eff}}{d z_0},
\label{eqofm}
\end{equation}
where the effective potential felt by the soliton is given by:
\begin{equation}
V_{\rm eff}(z)=\frac{1}{2}\Omega^2 z^2 + \frac{1}{2} V_0\left[1-\left(\frac{\pi^2}{3}-2\right)\frac{k^2}{6\mu}\right] \cos(2kz).
\label{Veff}
\end{equation}
At this point we should make the following remarks.
First, for the derivation of Eq.~(\ref{eqofm}), it was assumed that the background density of the condensate can be described via the TF 
approximation; this actually means that Eq.~(\ref{eqofm}) is valid only for sufficiently large number of atoms.
At the same time, however, the derivation of Eq.~(\ref{eqofm}) was done in the framework of the cubic NLS model, i.e., the weakly-interacting 
limit of Eq.~(\ref{1dDelgado}), which is valid for sufficiently small number of atoms. Nevertheless, Eq.~(\ref{eqofm}) may still be relevant to 
provide some estimates. It is known that the soliton oscillation frequency (for a BEC confined in a harmonic trap) is up-shifted for large number of atoms due to the dimensionality of the system (i.e., the effect of transverse dimensions, which are taken into regard in the derivation 
of Eq.~(\ref{1dDelgado})) \cite{Theo2007}. Particularly, one should expect that in the case of a purely harmonic trap the soliton oscillates with 
a frequency
%
$\omega_{\rm ds}>\Omega/\sqrt{2}$,
which can be found numerically in the absence of the optical lattice (note that $\Omega/\sqrt{2}$ is the characteristic value of 
the soliton oscillation frequency in a 1D harmonic trap of strength $\Omega$ 
in the TF limit~\cite{fms,mprizolas,muryshev,motion1,motion2,motion3,motion4,motion5,motion6}). For a more detailed discussion see Sec.~\ref{sec 4a}.
%
Moreover, for sufficiently small optical lattice wavenumbers (such as the chosen value
of $k=\pi/5.37 \mu m^{-1}$),
and large chemical potentials (TF limit)
the effective potential of Eq.(\ref{Veff}) can be simplified to the following form:
\begin{equation}
V_{\rm eff}(z)=
\omega_{\rm ds}^2 z^2 + \frac{1}{2} V_0 \cos(2kz).
\label{simpeffpo}
\end{equation}

We complete this Section by mentioning that the case of two (or more) spatially well separated solitons can also be treated analytically, taking into account the repulsive 
interaction potential between two solitons. In the case of two spatially separated solitons located at $ z_1$ and $z_2$, on top of a constant background with density 
$n_0$, the interaction potential takes the form \cite{draft6},
%
\begin{equation}
V^{\rm int}(z_1, z_2)=\frac{n_0}{\sinh^2(\sqrt{n_0}(z_2-z_1))}.
\end{equation}

%
%

\section{Bifurcation and BdG analysis}
\label{sec3}

Let us now proceed by investigating the existence, the stability and possible bifurcations of the lowest macroscopically excited states 
of Eq.~(\ref{1dDelgado}). We fix the parameter values as follows: 
$\Omega=0.1$, $k=\pi/5.37$ $\mu\text{m}^{-1}$ and $V_0=1.16 \times 10^{-12}$ eV; for such a choice,
the four lowest eigenvalues of the operator $1+\hat{H}$ (corresponding to the linear problem)
are found to be:
\begin{equation}
\omega_{0}=2.138 \times 10^{-12} {\rm eV}, \,\,
\omega_{1}= 2.142\times 10^{-12} {\rm eV}, \,\,
\omega_{2}=2.634\times 10^{-12} {\rm eV}, \,\,
\omega_{3}= 2.700\times 10^{-12} {\rm eV}, \,\,
\label{eigenvalues}
\end{equation}
%
and the energy of the first excited state is almost degenerate with that of the ground state, i.e., $\omega_{0}\sim\omega_{1}$.
\begin{figure}[htbp]
  \includegraphics[width=10cm,height=7cm]{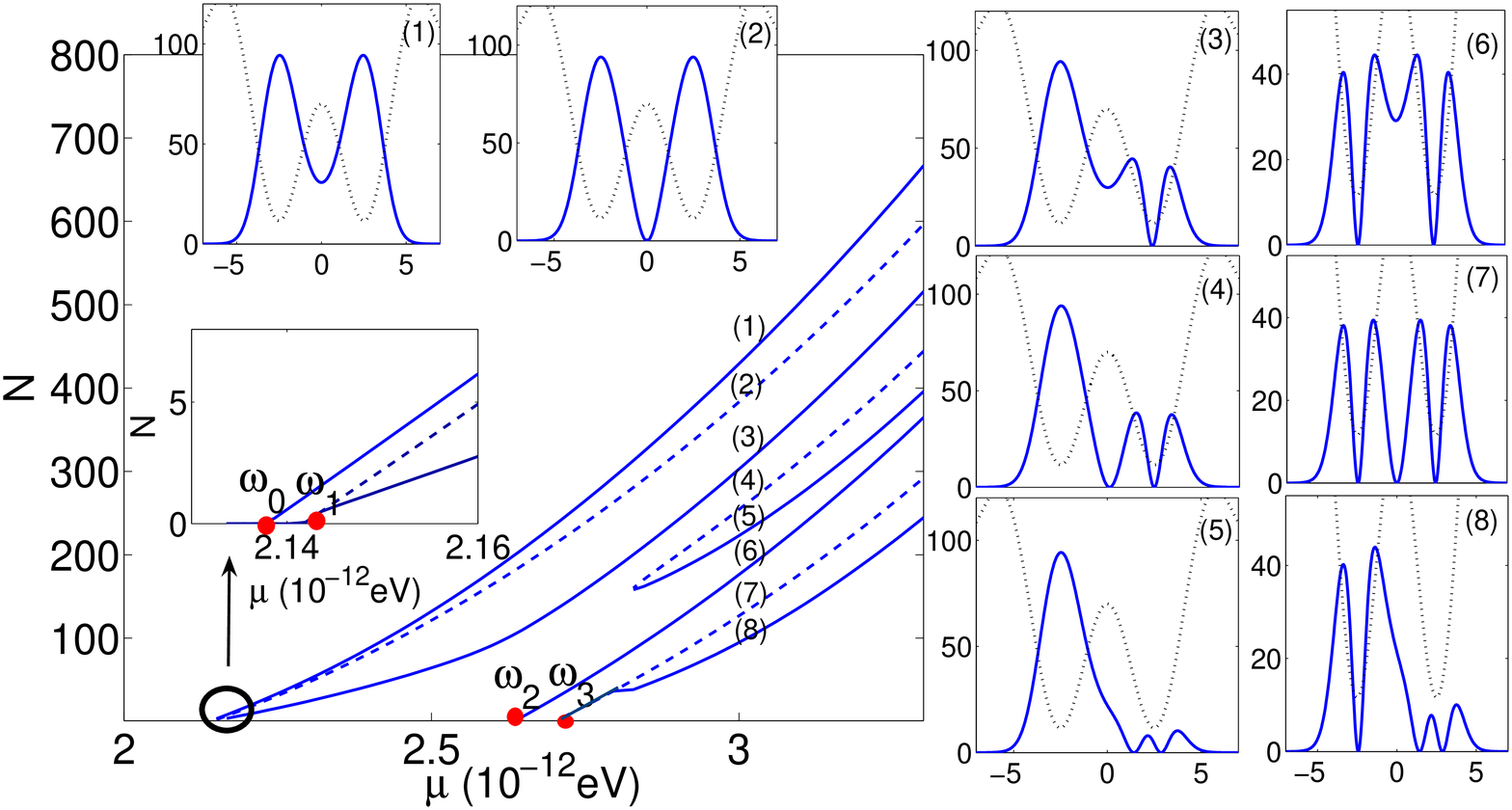}
  \caption{(Color online) Number of atoms as a function of the chemical potential for the different states. The potential parameters are
  $\Omega=0.1$, $k=\pi/5.37$ $\mu\text{m}^{-1}$ and $V_0=1.16 \times 10^{-12}$ eV.
  The insets show the densities of the different states for $\mu=3 \times 10^{-12}$ eV.}
\label{Fig: 1}
\end{figure}

Fig.~\ref{Fig: 1} shows the number of atoms as a function of the chemical potential for different branches of solutions. The insets show 
the spatial density profiles for $\mu=3\times 10^{-12}$ eV, with the units of the horizontal
and vertical axes being given by
$\mu m$ and
$1/\mu m$, respectively.

Branch (1) corresponds to the states with the largest number of atoms, for fixed chemical potential. The respective states are 
symmetric and have no nodes. Continuation of this symmetric branch to the linear limit ends at the eigenvalue $\omega_{0}$, corresponding to the 
ground state of the linear problem.

The second branch [branch (2)] starts from the first excited state in the linear limit, $\mu \rightarrow \omega_1$ as $N \rightarrow 0$.
The wavefunctions of the states of this branch are {\it antisymmetric} and have a node at the center of the barrier, while their densities are 
symmetric. From this density symmetric one-soliton branch, two asymmetric one-soliton branches, namely branch (3) and its mirror image with 
respect to the $z=0$ axis, bifurcate close to the linear limit, at $\mu \simeq 2.144 \times 10^{-12}$~eV -- see the close-up inset.
Let us define a local occupation number, i.e., number of atoms in the different wells, 
as the integral over the density up to the center of the barrier (see also Sec. V below). The occupation number in the well 
with the node is then smaller than the occupation number without the node.
This can be explained by
the fact that the state with one node is the first excited state, characterized by a higher energy than the one without a node, which is the 
ground state. Thus, in order to balance the chemical potential in both wells
one needs a larger interaction energy (i.e., more atoms) in the well without a node.
Note that the macroscopic quantum self-trapping (MQST) state as predicted in Ref \cite{smerzi} is not considered in what follows. 
The reason for that is that the MQST is a running phase state while our bifurcation analysis below focuses on the stationary states of the system.
Branches (4) and (5) have no linear counterparts since at $\mu \simeq 2.828\times10^{-12}$~eV they
``collide" and disappear. The states that belong to these branches are asymmetric, exhibiting two nodes. The state with larger number of atoms 
(for a fixed chemical potential) has one node approximately at the barrier and one node in one well. The other state has both nodes in one well. Once again, 
the occupation number in the well with two nodes is less than the one of the well with one node, so as to balance the chemical potential.

Branch (6) starts from the second excited state of the linear problem, hence
$\mu \rightarrow \omega_2$ as $N\rightarrow 0$. Therefore, the density of 
the respective state is symmetric and there is one node in each well.

Branch (7) starts from the third excited state of the linear problem, 
and $\mu\rightarrow \omega_3$ as $N\rightarrow 0$. The states belonging to 
this branch have three nodes, one located in each well and one at the barrier, and are symmetric. Close to the linear limit, at 
$\mu \simeq 2.795\times10^{-12}$~eV, two asymmetric three-soliton branches, namely branch (8) and its mirror image
with respect to the $z=0$ axis, bifurcate from this state. This state has two nodes in one well and one in the other. The occupation number in 
the well with more nodes is smaller than in the well with just one node in order to balance the chemical potential. Notice that there exist two 
more three-soliton branches without linear counterparts, but they only occur at higher chemical potentials --- where the BEC has occupied 
four-wells rather than two --- so they will not be considered here.

\begin{figure}[htbp]
\begin{minipage}[c]{8 cm}
\includegraphics[width=5.5cm,angle=270]{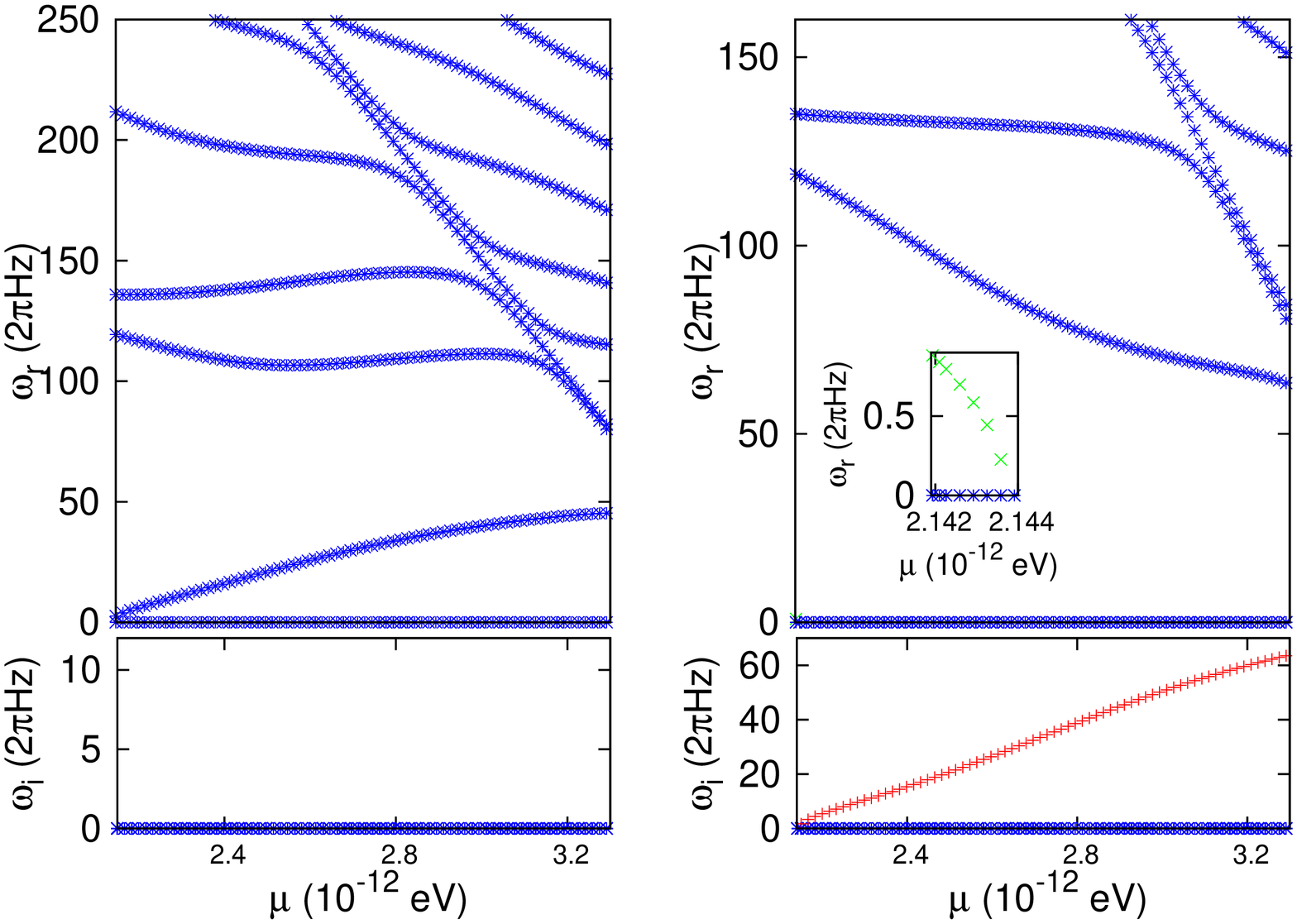}
  \end{minipage}
\begin{minipage}[c]{8 cm}
\includegraphics[width=5.5cm,angle=270]{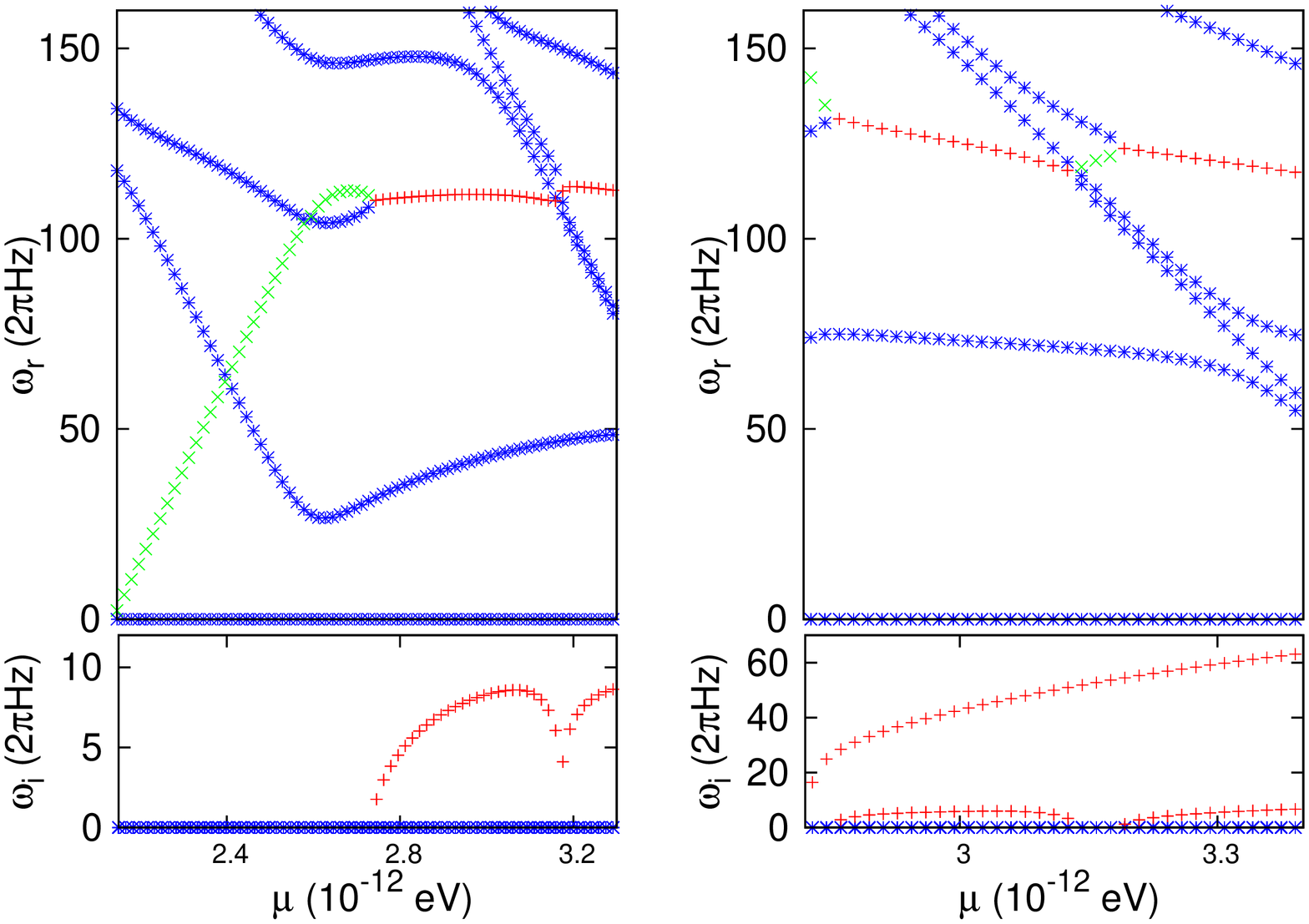}
  \end{minipage}
\begin{minipage}[c]{8 cm}
\includegraphics[width=5.5cm,angle=270]{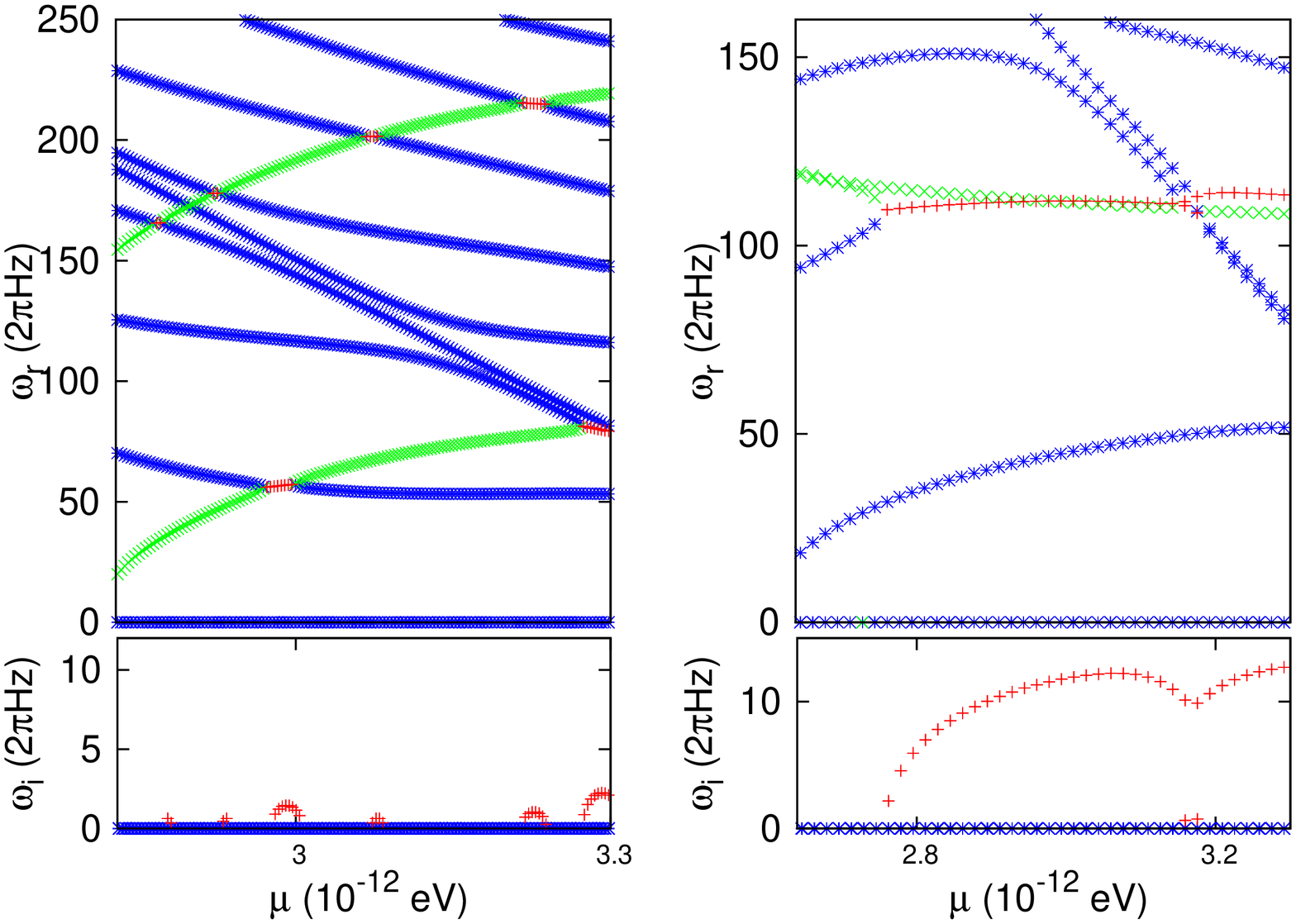}
  \end{minipage}
\begin{minipage}[c]{8 cm}
\includegraphics[width=5.5cm,angle=270]{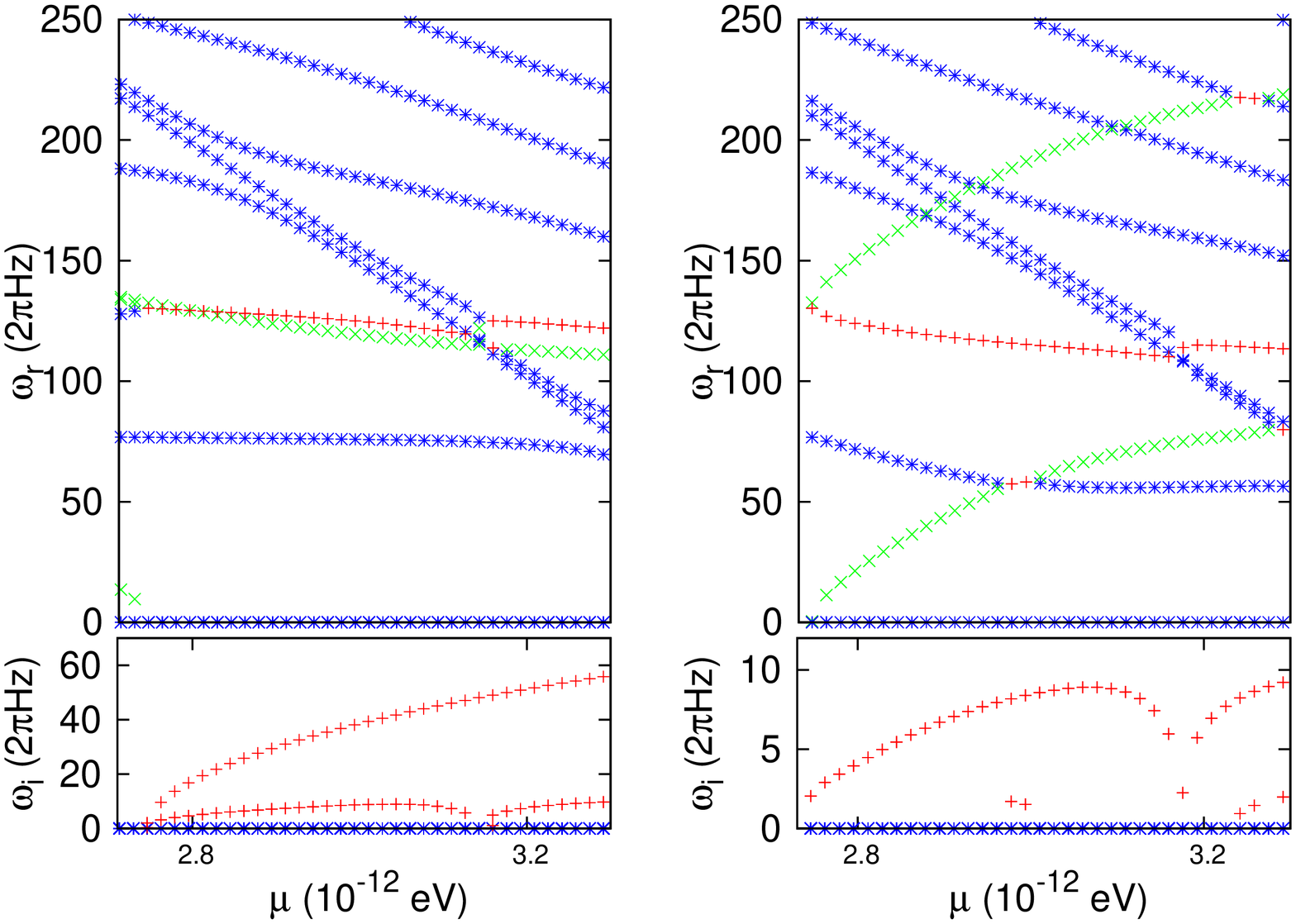}
  \end{minipage}
\caption{(Color online) BdG Spectrum of the states shown in Fig. \ref{Fig: 1}. First panel shows the spectrum of the first branch, second panel the spectrum of the second branch and so forth. The (blue) $\divideontimes$ symbol denotes an eigenmode of positive Krein sign (or zero Krein sign for a vanishing eigenfrequency), 
the (green) $\times$ symbol an eigenmode of negative Krein sign, and the (red) $+$ symbol an eigenmode associated to a vanishing Krein sign with complex/imaginary eigenfrequency.}
\label{Fig: 2}
\end{figure}

Next, let us study the  
stability of the states belonging to the above 
mentioned branches by considering the respective excitation spectra, 
shown in Fig.~\ref{Fig: 2}.
In this figure, the (blue) $\divideontimes$ symbol denotes a positive Krein 
sign mode (or a zero one for a vanishing frequency), the (green) $\times$ symbol a negative Krein sign mode, 
i.e., a negative energy {\it anomalous mode}, and the (red) ($+$) symbol a vanishing Krein sign, i.e., an eigenmode associated with complex/imaginary
eigenfrequency. Generally speaking, 
there are two different kinds of instability, corresponding to the cases of either a purely imaginary eigenfrequency, or a 
genuinely complex 
eigenfrequency. The latter case gives rise to the so-called oscillatory instability,
stemming from the collision of a negative energy mode with one of positive energy.

The first panel of Fig.~\ref{Fig: 2} shows the BdG spectra along the first branch. The imaginary part is zero for every value of the 
chemical potential, a fact reflecting the stability of this state. The absence of negative energy modes is expected, as this state is 
actually the ground state of the system. Since the BdG spectrum refers to all excitations of the state, one can attribute in the linear 
limit to each mode an excited state of the linear problem. The energy gap between the ground state and the first excited state corresponds 
to the energy of the first mode. Since the frequency of the so-called \textit{bosonic Josephson junction}, namely the oscillation frequency 
characterizing the transfer of atoms between the two wells, is equal to this frequency in the linear limit, this mode is connected to 
the oscillation of atoms between the two wells. We investigate this fact in more detail below (see
Sec.~\ref{section: Dynamics}) by direct simulations in the framework of
Eq.~(\ref{gerbier}). The frequency value of the second nonzero BdG mode of the ground state spectrum is correspondingly equal to the energy gap between the ground state and the second excited 
state in the linear limit, and so forth. We define the difference in the way that these frequency differences are positive.

The second panel shows the BdG spectra along the second branch, solutions of which represent states with a density minimum at the center of 
the barrier. In the linear limit one can assign to each mode an energy difference of the linear problem similar as in the BdG spectrum of the ground state. However in this case the frequency difference of the first excited state to the ground state is negative (according to our definition) leading to a negative energy mode. As one can see in the inset, the lowest mode has negative energy close to the linear limit.  However, increasing the chemical 
potential the eigenfrequency of this mode is moving rapidly towards the origin of the spectrum and, at $\mu= 2.1437\times10^{-12}$~eV, 
becomes and remains purely imaginary, thus carrying zero energy (red $+$ symbols). This happens exactly at the point where the third branch 
bifurcates as one can see in the inset of Fig. \ref{Fig: 1}. For the above mentioned potential parameters, this bifurcation happens in the weakly-interacting regime, where Eq.~(3) reduces 
to the cubic NLS equation. Thus, one can apply the Galerkin-type approach of Ref.~\cite{Theo06} and find that the critical value of the norm 
$N$ at which the bifurcation occurs is given by
\begin{equation}
N_\text{cr}=\frac{\omega_1-\omega_0}{3B-A_1},
\label{Ncr}
\end{equation}
where $\omega_{0,1}$ are the first two lowest eigenvalues of the operator $1+\hat{H}$, $A_1=\int{\psi_{1}^4dz}$, 
$B=\int{\psi_{0}^2\psi_{1}^2dz}$, while $\psi_{0,1}$ denote the ground state and the first excited state of the operator $1+\hat{H}$, 
respectively. One can also determine the critical chemical potential at which the symmetry breaking is expected to occur, 
namely $\mu_{cr}=1+\omega_{0}+3BN_{\text{cr}}$ \cite{Theo06} leading for (these parameter values) to 
$\mu_{\text{cr}}= 2.1436\times10^{-12}$~eV, which is in very good agreement with our numerical findings.


The third panel shows the BdG spectra along the third branch which bifurcates at
$\mu= 2.1437\times10^{-12}$~eV from the second one. All the eigenfrequencies of the arising asymmetric state are real close to the linear limit, 
thus it is concluded that this state is linearly stable and that the bifurcation is a supercritical pitchfork. This state has one negative 
energy mode as well, since it is a one-soliton state.
The eigenfrequency of the negative energy mode increases with increasing chemical potential,
passes through the mode
corresponding to the second excited state of the linear limiting case, and finally collides at some critical value of the chemical potential, $\mu = 2.76 \times 10^{-12}$~eV, 
with the mode
corresponding to the third excited state; this results in the generation of a quartet of unstable eigenfrequencies and, 
at this point, the state becomes oscillatory unstable. However, the magnitude of the instability is much smaller than in the previous case of 
the purely imaginary eigenfrequency.

The fourth and fifth panel show the spectra along the fourth and fifth branch, the solutions of which correspond to asymmetric two-soliton states 
having no linear counter-part. The previous one has got one imaginary and one negative energy mode and the latter one has two negative energy modes. These modes result from the presence of two dark solitons. The state considered 
in the fourth panel has two dark solitons, one soliton located approximately
at the barrier and one soliton in one well. The eigenfrequency corresponding to the former one is purely imaginary and looks similar to the mode of the symmetric one soliton state of the second panel. 
The eigenfrequency of the other anomalous mode decreases with increasing chemical potential and collides with a mode with positive energy, thus 
generating a quartet of complex eigenfrequencies. The magnitude of the imaginary part is again much smaller than the magnitude of the purely imaginary 
eigenfrequency.  The generation of dynamic instabilities is illustrated e.g. in panel five. 
The collision of one of the anomalous modes with a mode of positive Krein sign could lead to the emergence of a quartet of complex eigenfrequencies, resulting in the concurrent presence of an apparent
level crossing in the real part of the eigenfrequency along with a nonzero imaginary part of the relevant eigenfrequency mode.

At $\mu \simeq 2.828\times10^{-12}$~eV, the states of the fourth and fifth branch collide and disappear as can be seen in Fig. \ref{Fig: 1}. Just before the collision (for slightly larger $\mu$), the state of the 
fourth branch is unstable due to the presence of a mode with imaginary eigenfrequency in the BdG spectrum, while the state of the fifth branch 
is linearly stable since all the BdG eigenfrequencies have only real parts. Thus, it can be concluded that, at this critical point, a saddle-node 
bifurcation occurs, readily destroying these two states.

The sixth panel shows the BdG spectra along the sixth branch, the solutions 
of which correspond to symmetric two-soliton states. This branch 
has a linear counterpart, and we can assign to each mode in the linear limit an energy difference of the linear problem. The negative energy modes 
occur for negative energy differences, e.g., at the energy difference to the ground state and the first excited state. 
The anomalous modes here are close to being degenerate. Physically speaking, 
this reflects the fact that the dark solitons are 
essentially decoupled at the linear limit due to the presence of the optical lattice. The negative energy modes are not the 
modes with the lowest eigenfrequency.
The energy gaps between the second excited state and 
the third and fourth excited state, respectively, are smaller than the gaps between the 
ground state and the first excited state. Therefore, these modes with 
positive energy occur in the linear limit at lower frequencies. 
One of the negative energy modes collides at some critical value 
of the chemical 
potential with the mode stemming from the fourth excited state and forms a 
quartet of complex frequencies, thus leading to an oscillatory instability.

The seventh panel shows the BdG spectra along the seventh branch, 
the solutions of which corresponds to the symmetric 
three-soliton state exhibiting 
a linear counterpart. The negative energy modes occur at the energy gaps to the ground state, the first and the second excited states. The 
eigenfrequency of the latter becomes, for increasing $\mu$, purely imaginary at the point where the eighth branch bifurcates. The other two anomalous modes 
are almost degenerate, reflecting again the fact that the solitons in the 
different wells are almost decoupled. One of the negative energy modes 
collides with a mode with positive energy and forms a quartet of 
complex frequencies. However, the magnitude of the imaginary part is much 
smaller than the magnitude of the purely imaginary mode. The behavior of the purely imaginary mode is similar to the behavior of the mode in 
panel two (describing a single soliton at the center of the barrier). The behavior of the other two modes is similar to the modes in panel 
six describing one soliton in each well.

Finally, the eighth panel shows the BdG spectra along the eighth branch, the solutions of 
which corresponds to the asymmetric three-soliton state. 
This state, which has two solitons in one well and one soliton in the other well, emerges at $\mu\sim 2.74\times10^{-12}$~eV, where the seventh 
state (i.e., the one corresponding to the nonlinear continuation of the third excited state),
becomes unstable. The critical value of the number of atoms for which the bifurcation occurs can be predicted in the same way as for the bifurcation 
of the one soliton branch; the result is:
\begin{equation}
N_\text{cr}^{(3)}=\frac{\omega_3-\omega_2}{3D-C_1},
\label{Ncr2}
\end{equation}
where $\omega_{2,3}$ are the third and fourth lowest eigenvalues of operator $1+\hat{H}$, $C_1=\int{\psi_{3}^4dz}$, 
$B=\int{\psi_{2}^2\psi_{3}^2dz}$, while $\psi_{2,3}$ denote the second and third excited state of the operator $1+\hat{H}$, respectively. 
The corresponding critical chemical potential is then given by $\mu_{cr}^{(3)}=1+\omega_{2}+3BN_{\text{cr}}^{(3)}$ leading (for these parameters) 
to $\mu_{\text{cr}}^{(3)}= 2.7420\times10^{-12}$~eV, which is in very good agreement with the numerical critical point given above.

Right after the bifurcation, the asymmetric three soliton states are stable and, thus, we can conclude that the bifurcation
is of the supercritical pitchfork type. Two modes look similar to the negative energy modes in panel five describing a two-soliton state 
in one well and the third mode looks similar to a mode of one-soliton in a harmonic trap.

From the above analysis, we can readily derive some general conclusions concerning dark soliton states in a double well potential. The sum of 
negative energy modes and imaginary eigenfrequency modes 
is equal to the number of nodes in the wavefunction profile 
which, for large values of the chemical potential, represent 
dark solitons. In the linear limit, the BdG eigenfrequencies of a state 
correspond to the energy differences between that and other states of the 
linear system. 
For a negative energy difference the corresponding mode has a negative energy. 
Dark solitons located at the center of the barrier are known to 
be linearly unstable (for sufficiently high chemical potential) 
associated with a purely imaginary 
eigenfrequency \cite{weol,ichihara}. Thus, symmetric states with an 
odd number of nodes always 
have a mode which, above a critical value of the chemical potential, 
becomes purely imaginary. At this critical point additional stable 
asymmetric dark-soliton 
states emerge through supercritical pitchfork bifurcations.


\section{Statics vs. dynamics of dark soliton states}
\subsection{The one-soliton state} \label{sec 4a}


Having investigated in detail the statics of matter-wave dark solitons in double-well potentials, we now proceed to compare these results to the 
soliton dynamics, using the theoretical background exposed in Sec.~\ref{sec3}.B. Particularly, considering the case of a single dark soliton, 
we can readily obtain from Eqs.~(\ref{eqofm}) and (\ref{simpeffpo}) the following approximate equation of motion for the dark soliton center:
\begin{equation}
\frac{d^2 z_0}{dt^2} = -\omega_{\rm ds}^2 z_0 + \frac{1}{2}k V_0\sin(2kz_0).
\label{EOM}
\end{equation}
It is straightforward to find the fixed points $z_{\rm o}, p_{\rm o} \equiv dz_{\rm o}/dt$
of the above dynamical system, which are given by
%
\begin{equation}
p_{\rm o} =0, \qquad z_{\rm o}=\frac{kV_{0}}{2\omega_{\rm ds}^2}\sin(2kz_{\rm o}).
\label{fixpoint}
\end{equation}
%
It is clear that the fixed points correspond to intersections of the straight line
$f_1(z_{\rm o})=z_{\rm o}$ and the sinusoidal curve
$f_2(z_{\rm o}) =\frac{kV_{0}}{2\omega_{ds}^2}\sin(2kz_{\rm o})$.
The fixed point at $(z_{\rm o},p_{\rm o})=(0,0)$
exists for all choices of ($\Omega,V_{0},k$). This equilibrium corresponds to the symmetric one-soliton state, namely a stationary black 
soliton located at the center of the potential.
In the absence of the optical lattice (i.e., for $V_0=0$) this is the only existing fixed point, i.e., in the case of a purely harmonic trap, only a 
symmetric one-soliton state can exist, which originates from the first 
excited state of the linear problem. 
On the other hand, when the optical lattice is 
present (i.e., for $V_0 \ne 0$), 
this situation changes and more fixed points arise when the slope of the curve $f_2(z_{\rm o})$ at the origin becomes larger than the one of the 
straight line $f_1(z_{\rm o})$, namely when $k^2 V_{0}/\omega_{\rm ds}^2 > 1$, two new fixed points appear through a supercritical pitchfork, with the 
newly emerging states corresponding to the bifurcating 
asymmetric one-soliton states. Therefore, the bifurcating state appears only for a sufficiently strong optical lattice, such that 
$V_0 > V_{0,cr} \equiv \omega_{\rm ds}^2/k^2$. Note that for even larger values of the parameter $k^2 V_{0}/\omega_{\rm ds}^2$ more fixed 
points appear, corresponding to solitons located in further (i.e., more
remote from the trap center) wells of the optical lattice; for this reason,
these states are not considered herein.

Let us now investigate the spectral stability of the fixed points employing the linearization technique (see, e.g., Ref.~\cite{Strogatz}). 
Particularly, we first find the corresponding Jacobian matrix (evaluated at the fixed points), which is given by:
\begin{equation*}
{\bf J} = \left(
\begin{array}{cc}
0& 1  \\
-\omega_{\rm ds}^2+k^2V_{0}\cos(2kz_0) & 0\\
\end{array} \right).
\end{equation*}
When evaluated at the trivial fixed point $(z_{\rm o},p_{\rm o})=(0,0)$, the corresponding eigenvalue problem for the Jacobian,
$\det({\bf J}-\lambda{\bf I})=0$, leads to eigenvalues $\lambda^2=-\omega_{\rm ds}^2+k^2V_{0}$.
It is clear that in the case of $k^2 V_{0}/\omega_{\rm ds}^2<1$, the eigenvalues are imaginary (note that, accordingly, the eigenfrequencies 
are real since $\omega=i\lambda$) and, thus, the fixed point 
$(0,0)$ is a center.
In this case, if the dark soliton is weakly displaced from the center of the 
trap, it performs harmonic oscillations around the center
with an oscillation frequency given by $\omega_{\rm osc}=\sqrt{\omega^2_{\rm ds}-k^2V_{0}}$.
On the other hand, if $k^2\frac{V_{0}}{\omega_{\rm ds}^2}>1$ then the eigenvalues become real (and, accordingly, the eigenfrequencies 
are imaginary), hence
$(0,0)$ becomes a saddle point, while the respective
imaginary eigenfrequency is given by 
$\omega_{\rm osc}=i\sqrt{|\omega^2_{\rm ds}-k^2V_{0}|}$.
For the newly bifurcating (stable, since $V_{0}>\omega_{\rm ds}^2/k^2$) 
fixed points
%
one can obtain the following eigenfrequency
\begin{equation}
 \omega_{\rm osc}=\sqrt{\omega_{ds}^2-k^2V_{0} \cos(2kz_0)}.
\label{omega ODE}
\end{equation}
%
From the above analysis it is clear that  the
supercritical pitchfork occurs at
\begin{equation}
V_{0,cr}=\frac{\omega_{\rm ds}^2}{k^2}.
\label{V0cr}
\end{equation}
Based on the above, one can make the following estimates and
comparisons to numerical results:
in the TF limit, and in the weakly-interacting case 
(i.e., in the framework of the 1D cubic GP 
equation), the dark soliton oscillation frequency
is predicted to be $\omega_z/\sqrt{2}=28.28$Hz. However, as mentioned in
Sec.~\ref{sec3}.B, a larger value of this frequency is expected in the framework of Eq.~(\ref{gerbier}), due to the effect of the dimensionality of the system.
Employing a BdG analysis in the case of a pure harmonic trap, we find that the anomalous mode eigenfrequency 
(i.e., the soliton oscillation frequency) decreases with increasing chemical potential and asymptotically reaches the value 
$\omega_\text{ds}=30.36$Hz (see first panel of Fig. \ref{Fig: 3} below). Using this value (and for
$k=\pi/5.37 \mu m^{-1}$), one obtains $V_{0,cr}=9.6\times10^{-14}$~eV.

\begin{figure}[htbp]
\begin{center}
\includegraphics[width=8.5cm]{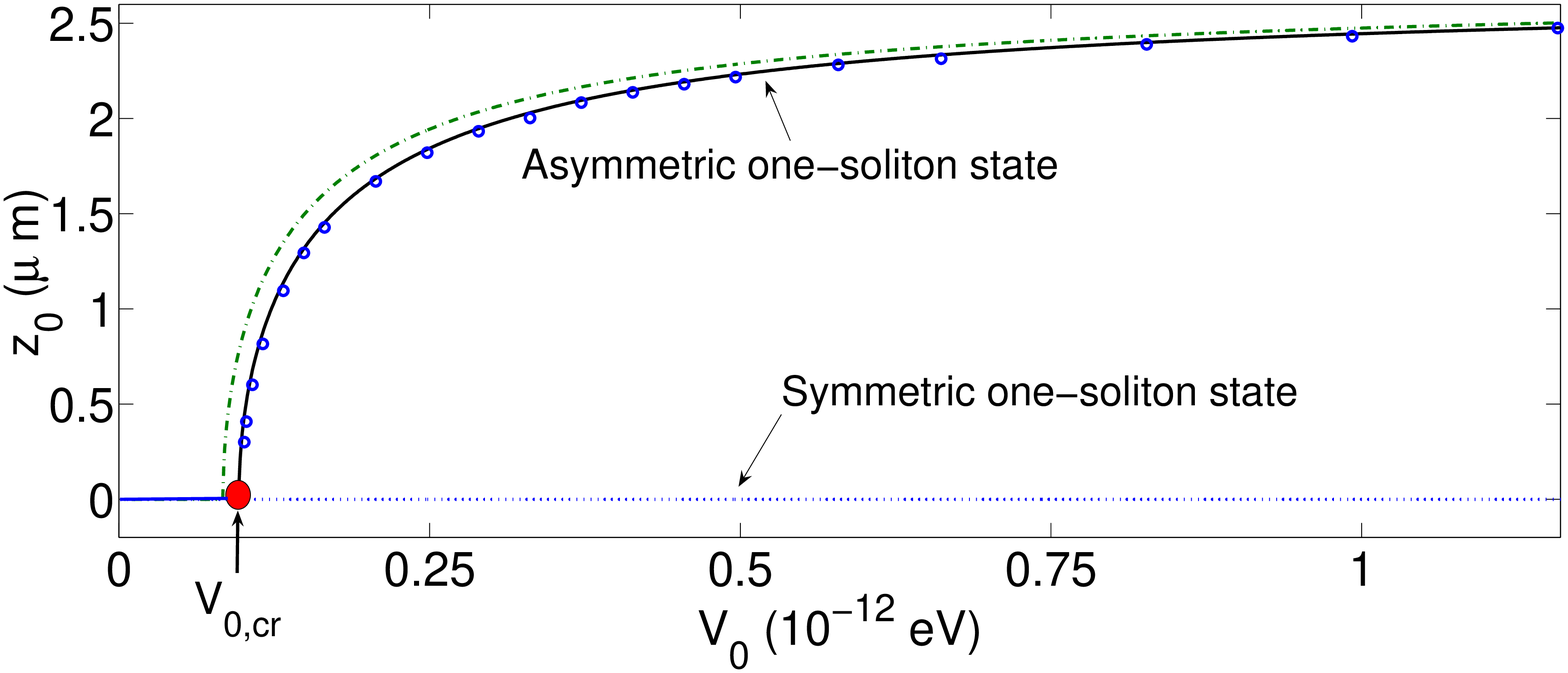}
\end{center}
\vspace{-0.5cm}
\includegraphics[width=6cm,angle=270]{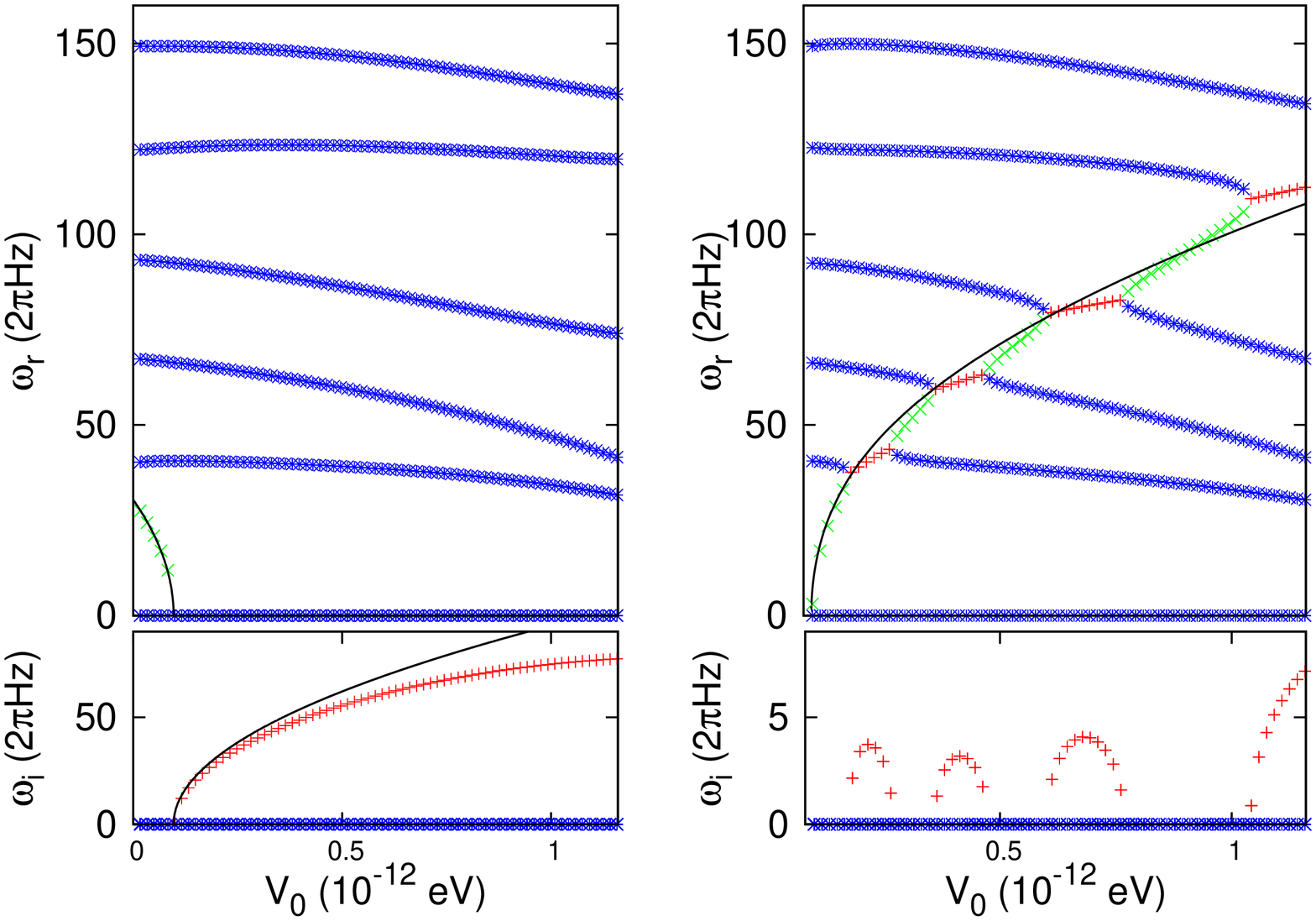}
\caption{(Color online) Top panel: Fixed points (lines) of the 
dynamical  system (\ref{EOM})  and position of the 
soliton (open circles) calculated by the Eq. (\ref{gerbier}) with $\mu=3.8 \times 10^{-12}$~eV. Bottom left (right) panel: 
BdG spectra of the symmetric (asymmetric) one-soliton state for $\mu=3.8 \times 10^{-12}$~eV as a function of $V_0$.}
\label{Fig: 4}
\end{figure}

In the top panel of Fig. \ref{Fig: 4}, we show the fixed points of the dynamical system (\ref{EOM})
as a function of the optical lattice strength $V_{0}$. The dash-dotted (green) line denotes the non-zero fixed point calculated from 
Eq.~(\ref{fixpoint}) with $\omega_{\rm ds}=\omega_{z}/\sqrt{2}$, while the solid (black) line denotes the non-zero fixed point 
calculated from Eq.~(\ref{fixpoint}) with $\omega_\text{\rm ds}=30.36$ Hz. Furthermore, the circles denote
the position of the node of the nonlinear stationary state --- i.e., the position of the dark soliton --- obtained by numerical integration 
of Eq.~(\ref{gerbier}) with $\mu=3.8 \times 10^{-12}$~eV (in the TF limit). Clearly, at $V_{0,cr}\simeq 9.6\times10^{-14}$~eV, the asymmetric 
one-soliton branch bifurcates through a supercritical pitchfork bifurcation; notice the excellent agreement between the two approaches.

The bottom left and right panels of Fig. \ref{Fig: 4} show the BdG spectra of the symmetric and asymmetric one soliton state, respectively, 
as a function of the optical lattice strength (for a fixed chemical potential, $\mu=3.8 \times 10^{-12}$~eV).
For sufficiently small lattice strength, $V_0<V_{0,cr}$, the symmetric state is stable, while when the lattice strength $V_0$ is increased, 
the magnitude of the instability grows and the prediction of
Eq.~(\ref{omega ODE}) deviates from the results obtained from the integration of Eq.~(\ref{1dDelgado}). To explain this discrepancy, it is relevant to recall that the result of the perturbation theory of Sec.~\ref{sec2}.B is valid only if
all perturbation terms are small and of the same order. Nevertheless, for the symmetric one soliton state --- with the soliton located at zero --- the 
magnitude of the perturbation term corresponding to the harmonic trap (optical lattice) has its smallest (largest) value; thus,
for large $V_0$, these
perturbations
cannot be of the same order.
As concerns the asymmetric state, the prediction of Eq.~(\ref{omega ODE}) is in fairly good agreement to the results obtained from 
the numerical integration of Eq.~(\ref{1dDelgado}). However, Eq.~(\ref{omega ODE}) cannot predict oscillatory instabilities (associated with 
the existence of complex eigenfrequencies), which
occur when the anomalous mode collides with a mode of the background condensate. This can readily be understood by the fact that the derivation 
of the equation of motion of the dark soliton relies on the assumption that the soliton is decoupled from the background (see, e.g., 
Refs.~\cite{motion2} and \cite{revnonlin} for details), which is not the case for oscillatory instabilities.

%



\begin{figure}[htbp]
\begin{minipage}[c]{8 cm}
\includegraphics[width=5.5cm,angle=270]{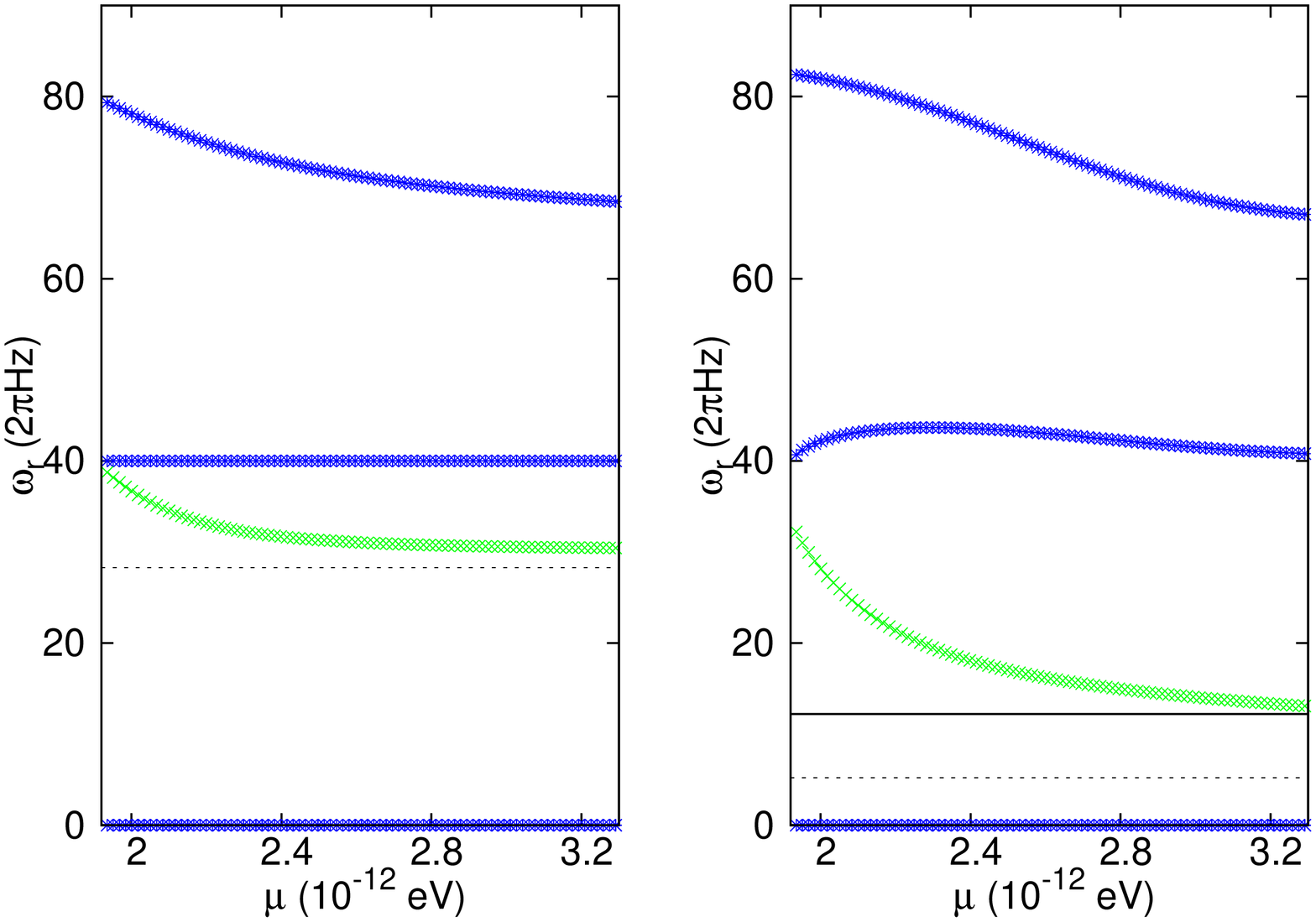}
  \end{minipage}
\begin{minipage}[c]{8 cm}
\includegraphics[width=5.5cm,angle=270]{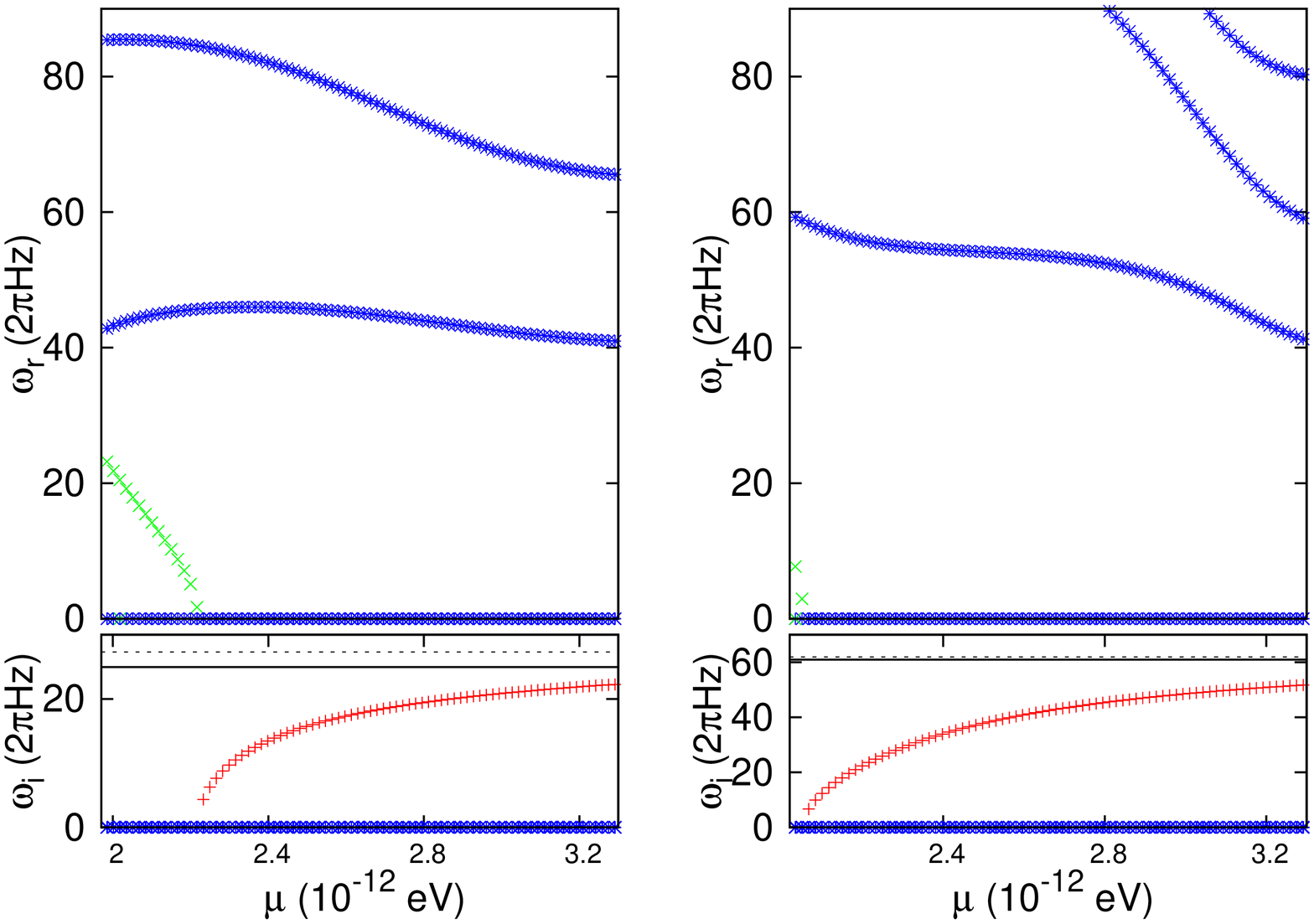}
  \end{minipage}
\caption{(Color online) BdG spectra for four different values of the optical lattice strength for the symmetric dark soliton state (with the soliton 
located at the trap center): 
from left to right, $V_0=0$, $V_0=8.3\times10^{-14}$ eV, $V_0=1.65\times10^{-13}$ eV and $V_0=4.96\times10^{-13}$ eV. 
The solid lines correspond to the predictions of Eq.~(\ref{omega ODE}). Symbol $\divideontimes$ (blue) denotes a positive (or zero for a vanishing eigenfrequency), symbol $\times$ (green) a negative and  symbol $+$ (red) a vanishing Krein sign mode (for non-zero eigenfrequencies).}
\label{Fig: 3}
\end{figure}
%

Next, let us study the continuation of both symmetric and asymmetric one-soliton states with increasing chemical potential for various values of 
the optical lattice strength.

First, we consider the symmetric one-soliton state and in
Fig.~\ref{Fig: 3} we show the BdG spectra along the corresponding branch.
The first panel shows the case of a pure harmonic trap, i.e., $V_0=0$, and the dashed line is the prediction from Eq.~(\ref{omega ODE}) 
using $\omega_{\rm ds}=\omega_{z}/\sqrt{2}$. It is clear that the eigenfrequencies are real, indicating the stability of this state for all values 
of the chemical potential. The figure shows the constant eigenfrequency of the dipolar mode [see straight (blue) line] at the value of the 
trap frequency $\omega_z$ and the anomalous mode bifurcating from the dipolar mode in the linear limit. As mentioned above, the eigenfrequency 
of the anomalous mode reaches asymptotically the value $\omega_\text{ds}=30.36$ Hz.

The second panel of Fig.~\ref{Fig: 3} shows the BdG spectrum for $V_0=8.3\times10^{-14}$~eV. All eigenfrequencies are real, so the state is still 
stable. The dashed line is the prediction by
Eq.~(\ref{omega ODE}) using $\omega_{ds}=\omega_{z}/\sqrt{2}$, while the 
solid line is the prediction with $\omega_\text{ds}=30.36$ Hz. 
The dipole mode remains approximately constant for such a small optical lattice strength. However, the degeneracy of the anomalous mode and the 
dipole mode in the linear limit disappears.

The third panel shows the spectra for $V_0 =1.65 \times10^{-13}$~eV. In this case, the anomalous mode becomes purely imaginary 
for increasing chemical potential, reflecting the fact that the state becomes unstable. Notice that the numerical result and the analytical prediction, 
$\omega_{ODE}=i\sqrt{|\omega^2_{\rm ds}-k^2V_{0}|}$, are in a good agreement for large values of the chemical potential, where the TF limit 
is reached and Eq.~(\ref{EOM}) is valid.
Since $V_0>{V}_{0,cr}$, this state is expected to be unstable in the TF limit.

The fourth panel shows the spectra for $V_0=4.96\times10^{-13}$~eV. In this 
case, the anomalous mode becomes purely imaginary already at small 
chemical potentials. Furthermore, the dipole mode eigenfrequency is now
significantly modified, upon the 
reported interval of changes of the chemical potential. 
The analytical prediction agrees again well with the 
numerical BdG result in the TF limit.

\begin{figure}[htbp]
\includegraphics[width=5.5cm,angle=270]{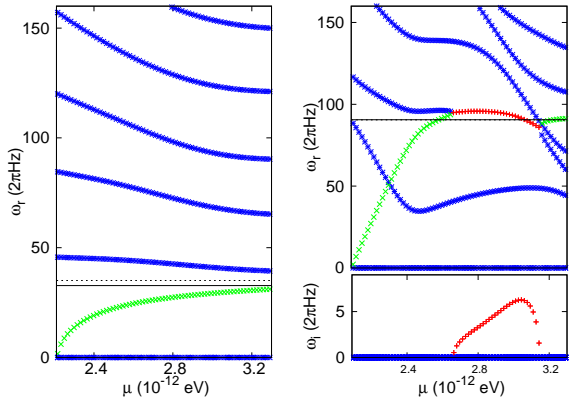}
\caption{(Color online) BdG spectra for $V_0 = 1.65\times 10^{-13}$~eV 
(left panel) and  $V_0=8.3\times10^{-13}$~eV (right panel) for the
asymmetric dark soliton state, bifurcating from the symmetric one. 
The notation of eigenstates is the same as in Fig. \ref{Fig: 3}.}
\label{Fig: 5}
\end{figure}

Next, we consider the continuation with increasing chemical potential of the asymmetric states bifurcating from the symmetric one when the latter 
becomes unstable.
Fig.~\ref{Fig: 5} shows the BdG spectra along the asymmetric one-soliton state for $V_0=1.65\times10^{-13}$~eV and $V_0=8.3\times10^{-13}$~eV. 
The analytical predictions of 
Eq.~(\ref{omega ODE}) with  $\omega_{\rm ds}=\omega_{z}/\sqrt{2}$ and $\omega_\text{ds}=30.36$Hz, respectively plotted with dashed and solid 
lines are
in good agreement
with the respective anomalous mode eigenfrequencies. In the case of $V_0=8.3\times10^{-13}$~eV, the difference between the result of 
Eq.~(\ref{omega ODE}) using $\omega_{ds}=\omega_{z}/\sqrt{2}$ and  $\omega_\text{ds}=30.36$ Hz is small, showing that the dynamics is 
mostly governed by the potential term stemming from the optical lattice.

\subsection{Two dark-soliton states}
We now consider the case of the two-soliton state, for which we need to incorporate the interaction potential in order to describe their dynamics. 
Particularly, according to the theoretical framework of Sec.~\ref{sec2}.B,
the effective potential describing the dynamics of two solitons located at $z_1$ and $z_2$ has the following form:
\begin{equation}
V_{\text{eff}}(z_1,z_2)=
\omega_{\rm ds}^2 (z_1^2+z_2^2) + \frac{V_0}{2}(\cos(2kz_1) +  \cos(2kz_2)) +\frac{2n_0}{\sinh^2(\sqrt{n_0}(z_2-z_1))}.
\label{eff. pot. two solitons}
\end{equation}
%
Using $d^2 z_i/dt^2 = -(1/2) dV_\text{eff}(z_1,z_2)/dz_i$ (with $i=1,2$) as per Eq.~(\ref{eqofm}), it is straightforward to derive
the following system of equations of motion,
\begin{eqnarray}
\frac{d^2 z_1}{dt^2} &=& -\omega_{\rm ds}^2 z_1 + \frac{1}{2}k V_0\sin(2kz_1) -2n_0^{3/2}\coth(\sqrt{n_0}(z_2-z_1)){\rm csch}^2(\sqrt{n_0}(z_2-z_1)),
\label{EOM two soliton1} \\
\frac{d^2 z_2}{dt^2} &=& -\omega_{\rm ds}^2 z_2 + \frac{1}{2}k V_0\sin(2kz_2) +2n_0^{3/2}\coth(\sqrt{n_0}(z_2-z_1)){\rm csch}^2(\sqrt{n_0}(z_2-z_1)),
\label{EOM two soliton2}
\end{eqnarray}
which, accordingly, leads
to the following system
for the fixed points ($z^{(1)}_{\rm o}$, $z^{(2)}_{\rm o}$):
\begin{eqnarray}
z^{(1)}_{\rm o} &=&
\frac{V_0 k}{2\omega_{\rm ds}^2}\sin(2kz^{(1)}_{\rm o})-\frac{n_0^{3/2}}{\omega_{\rm ds}^2} \coth(\sqrt{n_0}(z^{(2)}_{\rm o}-z^{(1)}_{\rm o})){\rm csch}^2
(\sqrt{n_0}(z^{(2)}_{\rm o}-z^{(1)}_{\rm o})),
\label{fixpoint two sol1}\\
z^{(2)}_{\rm o} &=&
\frac{V_0 k}{2\omega_{\rm ds}^2}\sin(2kz^{(2)}_{\rm o})+\frac{n_0^{3/2}}{\omega_{\rm ds}^2} \coth(\sqrt{n_0}(z^{(2)}_{\rm o}-z^{(1)}_{\rm o})){\rm csch}^2
(\sqrt{n_0}(z^{(2)}_{\rm o}-z^{(1)}_{\rm o})).
\label{fixpoint two sol2}
\end{eqnarray}
Thus, the fixed points  for the two soliton states depend on the density of the background cloud and thereby on the chemical potential.
Due to the presence of the external potentials, the density of the background is inhomogeneous. In our analysis, we replace the inhomogeneous 
density, with the density of the background at $z=0$ using the TF approximation; namely, $n_0\rightarrow n_0(z=0)=((\mu-V_0)^2-1)/4$.

Considering small deviations from the fixed points, i.e.,
$z_i=z^{(i)}_{\rm o}+\epsilon \eta_i \exp(i\omega t)$ (with $i=1,2$), leads to the eigenvalue problem for the normal modes:
\begin{equation}
\omega^2\mathbf\eta = \left(
\begin{array}{cc}
A(z^{(1)}_{\rm o})+B(z^{(1)}_{\rm o},z^{(2)}_{\rm o})& -B(z^{(1)}_{\rm o},z^{(2)}_{\rm o})
\\
-B(z^{(1)}_{\rm o},z^{(2)}_{\rm o}) & A(z^{(2)}_{\rm o})+B(z^{(2)}_{\rm o},z^{(1)}_{\rm o})
\\
\end{array} \right)\mathbf{\eta}
\label{2solitonfreqs}
\end{equation}
with
\begin{eqnarray}
A(x)&=&\omega_{\rm ds}^2-V_0 k^2\cos(2kx) \\
B(x,y)&=& 4 n_0^2 \coth^2(\sqrt{n_0} (x - y)) {\rm csch}^2(\sqrt{n_0} (x - y)) +
  2 n_0^2 {\rm csch}^4(\sqrt{n_0} (x - y)).
\end{eqnarray}
Diagonalizing the above matrix, we obtain the eigenfrequencies of the two 
anomalous modes of the two-soliton states.
Notice that since different two soliton states lead to
different fixed points,
the oscillation frequencies for the two solitons will differ as well. 
Negative eigenvalues of this $2 \times 2$ system lead to imaginary frequencies describing unstable states; 
however, as mentioned in the previous section, 
one cannot describe oscillatory instabilities.

\begin{figure}[htbp]
\begin{center}
\includegraphics[width=8.5cm]{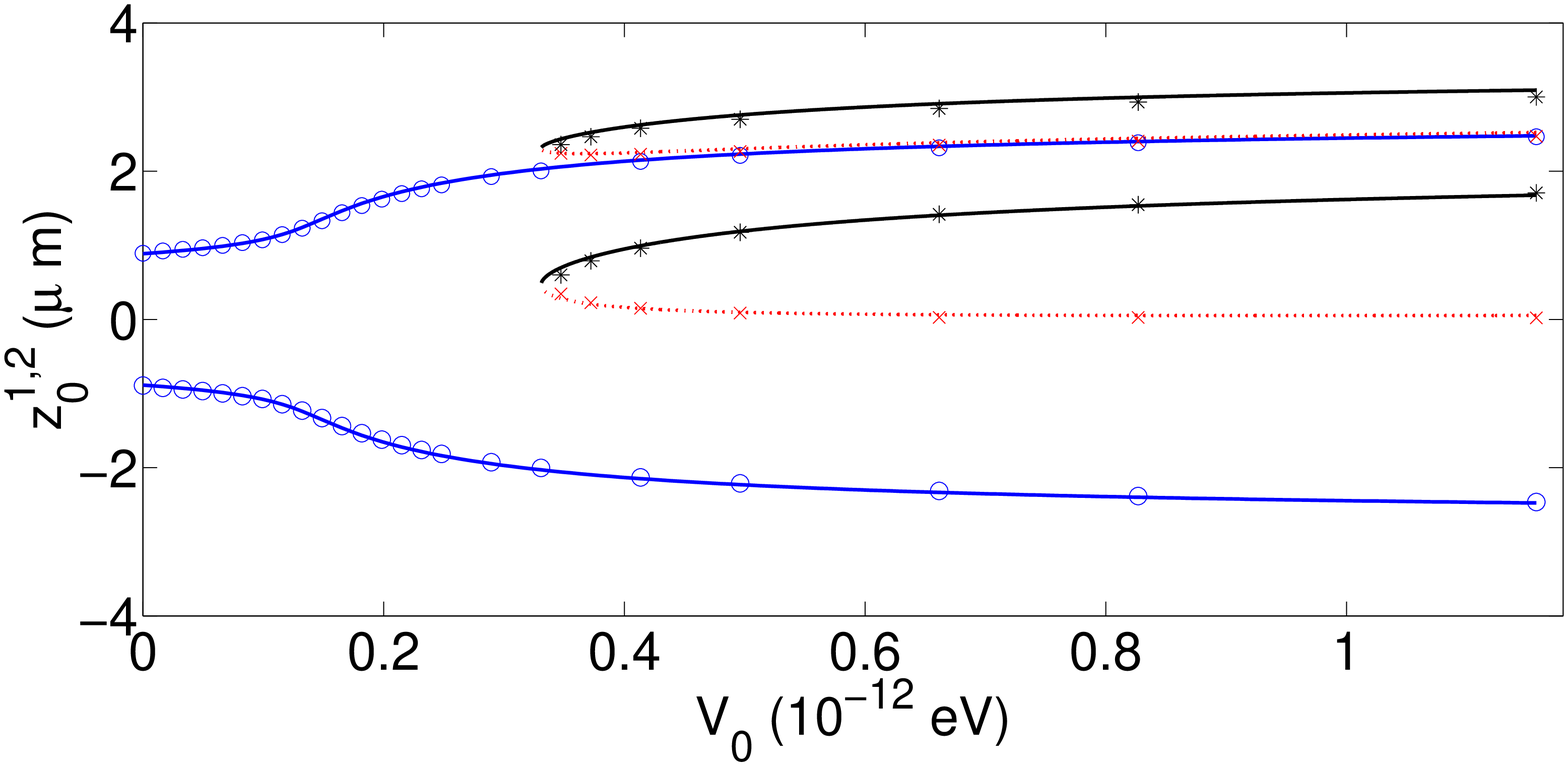}
\end{center}
\vspace{-0.5cm}
\includegraphics[width=6cm,angle=270]{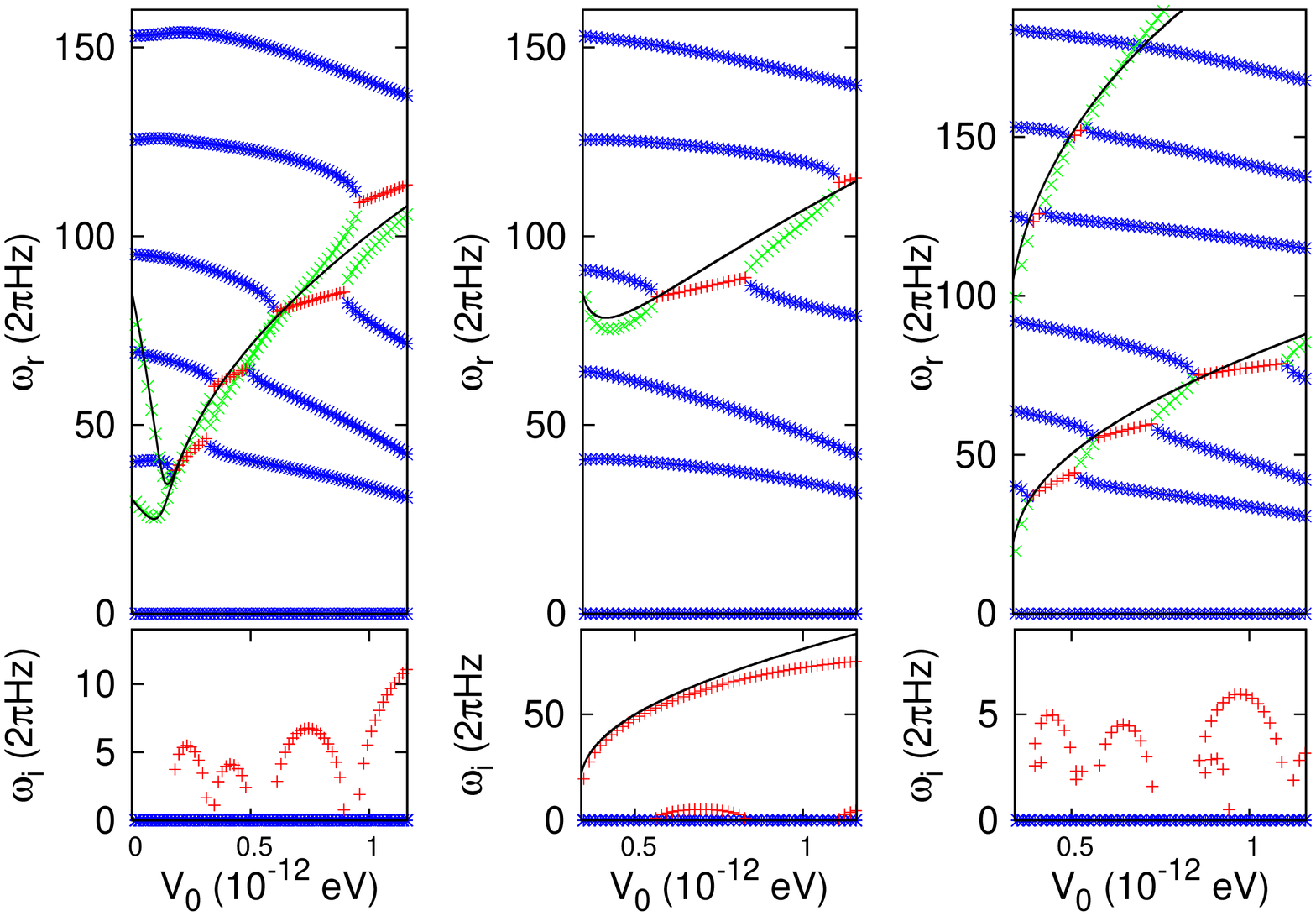}
\caption{(Color online) Top panels: Fixed points (lines) of the system 
(\ref{fixpoint two sol1})-(\ref{fixpoint two sol2})
and position of the solitons (circles for the symmetric two-soliton state,
$\times$ for the two-soliton branch with one soliton near the center
and one in one well, and stars for the state with both dark solitons
in the same well) 
calculated by the numerical integration of Eq.~(\ref{gerbier})
for $\mu=3.8 \times 10^{-12}$~eV.
Bottom panels:  BdG spectra of the two-soliton states for $\mu=3.8 \times 10^{-12}$~eV as a function of $V_0$; left corresponds to the symmetric two-soliton
state, middle to the always unstable state with one soliton near the center
and one in one well, while the right corresponds to the state 
with both dark solitons in the same well. Notice also the agreement of
the theoretically predicted modes from Eq. 
(\ref{2solitonfreqs}) with the numerical ones.}
\label{Fig: 6}
\end{figure}

In the top panel of Fig.~\ref{Fig: 6}, we show the fixed points of the system (\ref{fixpoint two sol1})-(\ref{fixpoint two sol2}) and the zero points of the solutions with 
two nodes of Eq.~(\ref{gerbier}) corresponding to the positions of the two solitons as a function of $V_0$ (the chemical potential is fixed: 
$\mu=3.6 \times 10^{-13}$~eV). For small $V_0$ there exists only the 
symmetric solution with two nodes. 
However, for  $V_{0,cr}=3.4 \times 10^{-12}$~eV, 
two asymmetric states appear stemming from a saddle-node bifurcation. For all three states, the agreement between the fixed points obtained from
Eqs.~(\ref{fixpoint two sol1})-(\ref{fixpoint two sol2}) and the position of the solitons is excellent.

The bottom panels of Fig.~\ref{Fig: 6} show the BdG spectra for the symmetric and the two asymmetric two soliton states as  a function of $V_0$ 
for fixed $\mu=3.8 \times 10^{-12}$~eV. The solid lines correspond to the normal modes of Eqs.~(\ref{EOM two soliton1})-(\ref{EOM two soliton2}) 
around the corresponding fixed points. For a small optical lattice strength $V_0$, the two anomalous modes of the symmetric states have a different 
magnitude, which indicates that the solitons are coupled. However, for increasing $V_0$, this difference decreases and, thus, the coupling between 
the soliton becomes weaker. Notice that for an optical lattice strength $V_0 \simeq 3.6\times 10^{-12}$ the two modes become of the same order, a fact 
that indicates that the two solitons become eventually decoupled. 
The prediction of Eqs.~(\ref{EOM two soliton1})-(\ref{EOM two soliton2}) agrees very well with the numerical results obtained via the 
BdG analysis for the anomalous modes. 
The second panel shows the BdG spectrum of the asymmetric state with one soliton located approximately at the center of the barrier. The latter, leads to 
a purely imaginary mode with an increasing magnitude with increasing $V_0$. The increase of the magnitude of this mode results from the increase 
of the height of the barrier. For large $V_0$ the prediction of Eqs.~(\ref{EOM two soliton1})-(\ref{EOM two soliton2}) deviates from the result 
obtained in the framework of Eq.~(\ref{gerbier}): the contributions of the perturbation terms are not of the same order and, thus, perturbation 
theory fails in this case. The third panel shows the BdG spectrum for the second asymmetric state with two solitons in one well. The difference 
between the modes is always large, a fact indicating the strong coupling between the two solitons. With increasing lattice strength $V_0$, the 
eigenfrequencies of the anomalous modes increases too. The predictions of Eqs.~(\ref{EOM two soliton1})-(\ref{EOM two soliton2}) agree once again well with the BdG results; nevertheless, as in previous cases, Eqs.~(\ref{EOM two soliton1})-(\ref{EOM two soliton2}) cannot 
predict oscillatory instabilities.

\begin{figure}[htbp]
\begin{minipage}[c]{8 cm}
\includegraphics[width=5.5cm,angle=270]{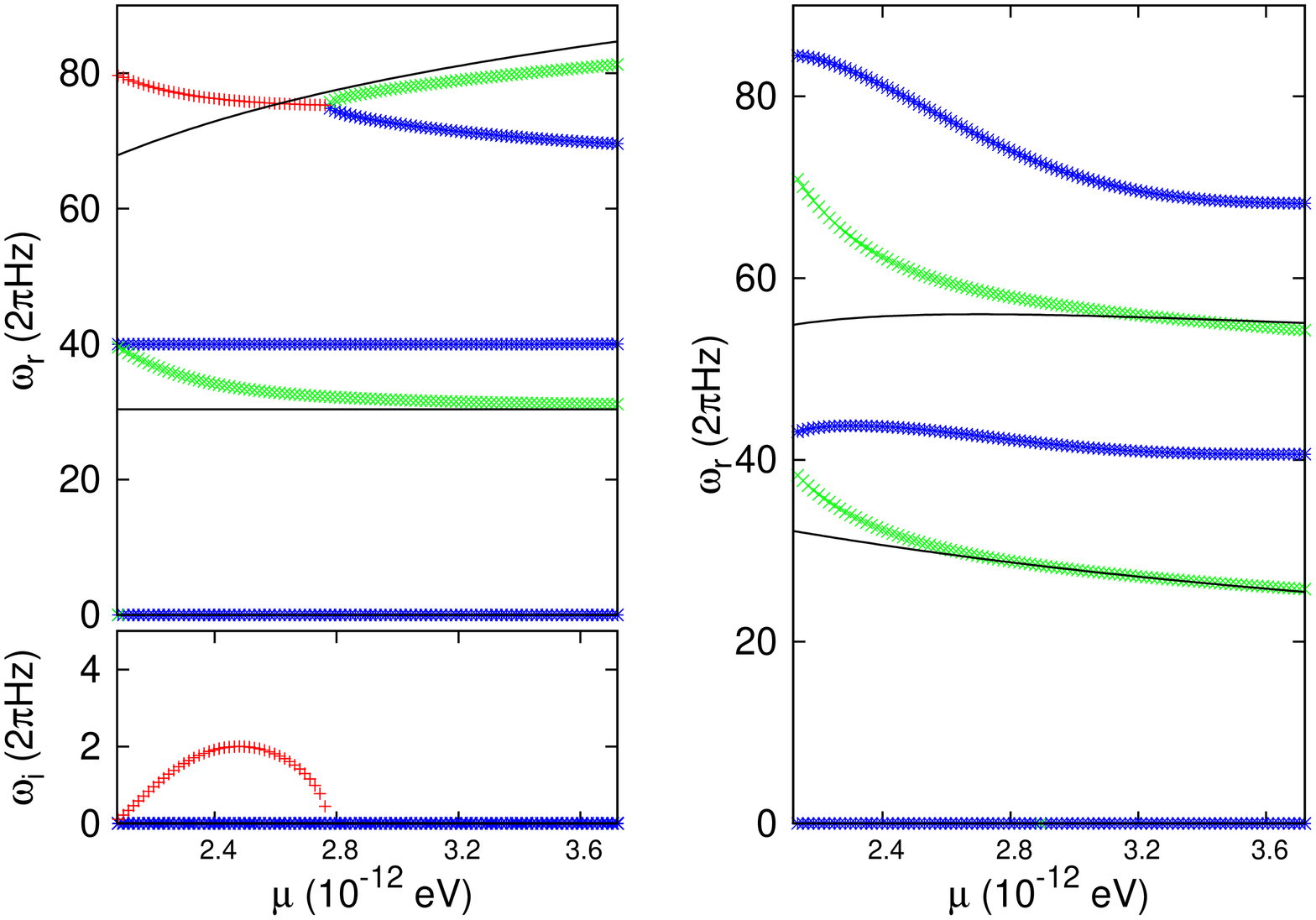}
  \end{minipage}
\begin{minipage}[c]{8 cm}
\includegraphics[width=5.5cm,angle=270]{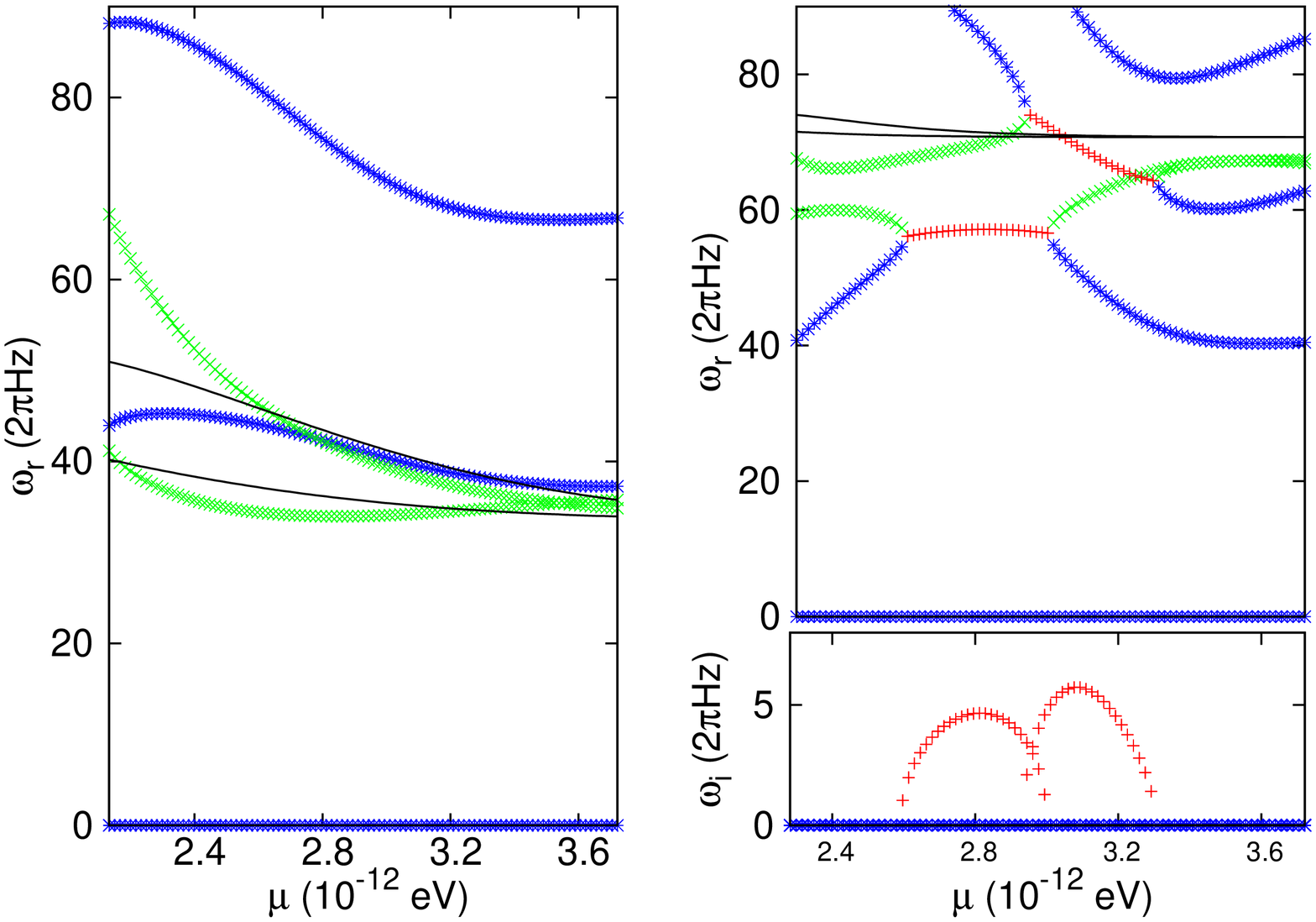}
  \end{minipage}
\caption{(Color online) BdG spectra for different values of the optical lattice strength; from left to right, 
$V_0=0$, $V_0=8.3\times10^{-14}$ eV, $V_0=1.65\times10^{-13}$ eV and $V_0=4.96\times10^{-13}$ eV. 
The 
solid lines are the predictions of Eqs.~(\ref{EOM two soliton1})-(\ref{EOM two soliton2}) with a numerically determined oscillation frequency ($\omega_{\rm ds}$).}
\label{Fig: 7}
\end{figure}

Next, let us consider a continuation with respect to the chemical potential.
Fig. \ref{Fig: 7} shows the BdG spectra, and the predicted oscillation frequencies, for the symmetric two soliton state for different values 
of the optical lattice strength: 
$V_0=0$, $V_0=8.3\times10^{-14}$ eV, $V_0=1.65\times10^{-13}$ eV and $V_0=4.96\times10^{-13}$ eV. 
In the case of a purely harmonic potential, $V_0=0$, one observes the constant eigenfrequency of the dipolar mode at $\omega_z$, as well as the eigenfrequency of one anomalous mode bifurcating from the dipolar mode in the linear limit. The eigenfrequency at $2\omega_z$ is imaginary due to a degeneracy of a negative and a positive Krein sign mode. The former one corresponds to the second anomalous mode and the latter one to the quadrupole mode. This degeneracy is lifted for increasing chemical potential. The predictions of Eqs.~(\ref{EOM two soliton1})-(\ref{EOM two soliton2}) agree very well with the BdG results for the anomalous modes in the case of large chemical potentials.

For $V_0=8.3\times10^{-14}$ eV the degeneracy between the anomalous modes and the dipolar and the quadrupole mode is lifted in the linear limit. 
Thus, there exists an energy gap between the anomalous mode and the dipolar mode and, more importantly, an energy gap between the second anomalous 
mode and the quadrupole mode. A particularly useful conclusion stemming from the above findings is that the two soliton state can be stabilized in 
a harmonic trap by adding a small optical lattice.

For $V_0=1.65\times10^{-13}$ eV the system is stable as well. The anomalous modes decrease with increasing chemical potential and reach similar values 
for a large chemical potential. This denotes that the two solitons are decoupled for a large chemical potential. This can be understood at least in the 
framework of the effective potential (\ref{eff. pot. two solitons}): the range of the interaction potential scales with the density $n_0$ of the 
background atom cloud 
and thereby with the chemical potential as well. Thus, for a large chemical potential, i.e., a large density, the interaction between solitons is strong but of short range too.

For $V_0=4.96\times10^{-13}$ eV the anomalous modes have similar 
values even in the linear limit. Thus, 
the two solitons are already almost decoupled 
in the linear limit due to their large separation. The two anomalous modes collide with the dipolar and quadrupole modes for increasing chemical 
potential, leading to oscillatory instabilities, and bifurcate from these modes again for a larger chemical potential. Although the predictions of 
Eqs.~(\ref{EOM two soliton1})-(\ref{EOM two soliton2}) agree qualitatively with the BdG results, there is no quantitative agreement due to the 
crossings and collisions of the soliton modes with the modes of the condensate.


\section{Dark soliton dynamics: numerical results} 
\label{section: Dynamics}

So far, we have examined the existence, stability and bifurcations of the 
branches of dark-solitonic states in the double-well potential setting. 
We now proceed to investigate the dynamics of these states by 
direct numerical integration of Eq. (\ref{gerbier}), 
using as initial conditions stationary soliton states, perturbed along the 
direction of eigenvectors associated with particular eigenfrequencies
(especially ones associated with instabilities).   
To better explain the above, we should recall that anomalous modes are 
connected to the positions and oscillation frequencies of the solitons; 
therefore, perturbations of stationary dark soliton states along the 
directions of the eigenvectors of the anomalous modes 
result in the displacement of the positions of the solitons, thus giving rise 
to interesting soliton dynamics.
%

\begin{figure}[htbp]
\includegraphics[width=5.5cm,angle=270]{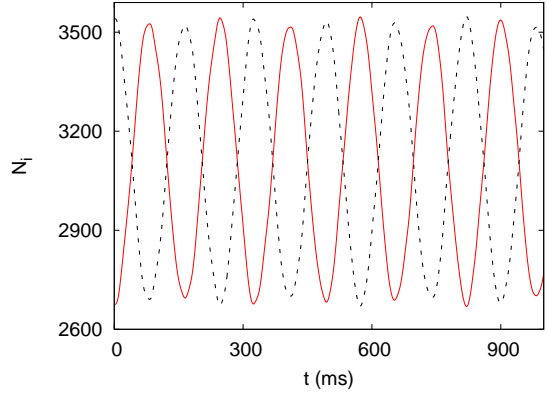}
\caption{(Color online) Time evolution of the number of atoms $N_\text{L}$ (solid line) 
and $N_\text{R}$ (dashed line) in the left and right well of the double-well potential, respectively, 
for the ground state of the system perturbed by the lowest BdG mode. Parameter values are  
$\mu=3 \times 10^{-12}$ eV and $V_0 =1.12 \times 10^{-12}$ eV.}
\label{Fig: 8a}
\end{figure}

Before proceeding with the dynamics of the (excited) soliton states, let us make a remark concerning the ground state of the system. 
As mentioned above, the value of the lowest BdG mode of the ground state is given, in the linear limit, by the energy difference 
between the ground state and the first excited state. However, this energy gap determines the oscillation frequency of atoms between 
the wells of the double-well potential, 
which suggests that the lowest BdG mode is connected to an oscillation between the wells.
Let us define the atom numbers $N_{\rm L}$ and $N_{\rm R}$ in the left and right well, respectively, as  
$N_{\rm L}=\int_{-\infty}^0 |\psi|^2 dz$ and $N_{\rm R}=\int_{0}^{+\infty} |\psi|^2 dz$  
%
%
(the total number of atoms is $N=N_\text{L}+N_\text{R}$). In Fig \ref{Fig: 8a}, we show the time evolution of $N_\text{L}$ (solid line) 
and $N_\text{R}$ (dashed line) for the ground state, when perturbed by the eigenvector corresponding to the lowest BdG mode (parameter values are   
$\mu=3 \times 10^{-12}$ eV and $V_0=1.12 \times 10^{-12}$ eV). One can clearly observe the oscillation of the atom numbers in the different wells, with  
a characteristic frequency known as Josephson oscillation 
frequency~\cite{smerzi}, determined by the magnitude of the 
respective BdG eigenmode. 

\begin{figure}[htbp]
\includegraphics[width=5.5cm,angle=270]{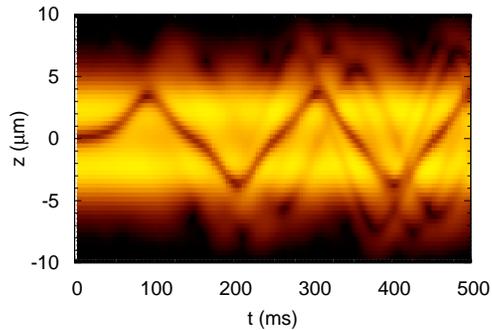}
\caption{(Color online) Contour plot showing the space-time 
evolution of the density of the condensate, which carries the symmetric 
one-soliton state; dark corresponds to minimal and white to maximal density. 
The soliton is unstable and departs from the trap center after a short time period, 
performing subsequently large amplitude oscillations. 
Parameter values are $\mu=3 \times 10^{-12}$ eV and $V_0=4.96 \times 10^{-13}$.}
\label{Fig: 8}
\end{figure}

Let us now proceed with the one-soliton states. 
Fig.~\ref{Fig: 8} shows the time evolution of the symmetric one soliton state perturbed by the eigenvector corresponding to 
the unstable mode, for $\mu=3 \times 10^{-12}$ eV and $V_0=4.96 \times 10^{-13}$ eV. The instability manifests itself as a drift of the soliton from the trap center to the rims of the background atom cloud, where 
the soliton is reflected back and forth. One clearly observes that the soliton is located at an unstable position, since a small perturbation leads 
to a large amplitude oscillation of the soliton.
\begin{figure}[htbp]
\includegraphics[width=5.5cm,angle=270]{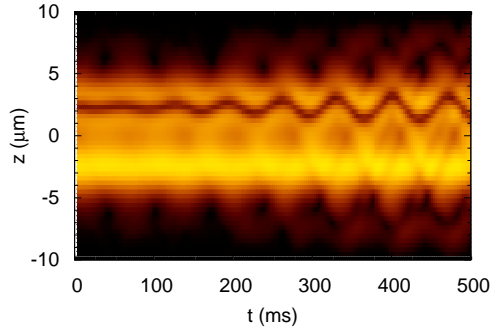}
\caption{(Color online) Same as in Fig.~\ref{Fig: 8}, but for the asymmetric 
one-soliton state, 
for $\mu=3 \times 10^{-12}$ eV for $V_0=8.27 \times 10^{-13}$ eV.}
\label{Fig: 9}
\end{figure}
On the other hand, Fig.~\ref{Fig: 9} shows the time evolution of the asymmetric one soliton state perturbed by the eigenvector corresponding to the
unstable mode for $\mu=3 \times 10^{-12}$ eV for 
$V_0=8.27 \times 10^{-13}$ eV. The corresponding state is subject to an oscillatory instability, which  
results in a growing amplitude of the soliton oscillation. However, in comparison to the previous case, the magnitude of the instability is much smaller 
and, thus, the increase of the amplitude is smaller (for the same time scale) as well.

%

Next, we consider the multi-soliton states and start with the 
symmetric two-soliton state which, for 
$\mu=3.3 \times 10^{-12}$ eV and $V_0=8.27 \times 10^{-14}$ eV, is found to be linearly stable. 
Numerical simulations, using as initial conditions this state perturbed by the eigenvectors corresponding to both anomalous modes, 
reveal that the two solitons perform oscillations with a constant amplitude and different frequencies. 
Specifically, the mode with smaller amplitude performs an in-phase oscillation (with the two solitons moving towards the same direction), 
whereas the mode with larger amplitude performs an out-of-phase oscillation (with the two solitons moving towards opposite directions) 
with a larger oscillation frequency. The existence of two different oscillation frequencies indicates the coupling of the dark solitons.
This situation changes for a large optical lattice strength (e.g., $V_0=1.65 \times 10^{-13}$ eV): in this case, the two 
dark solitons are decoupled, since the numerical simulations show that both the in-phase and out-of phase oscillations are characterized 
by almost the same frequency.

%
\begin{figure}[htbp]
\includegraphics[width=5.5cm,angle=270]{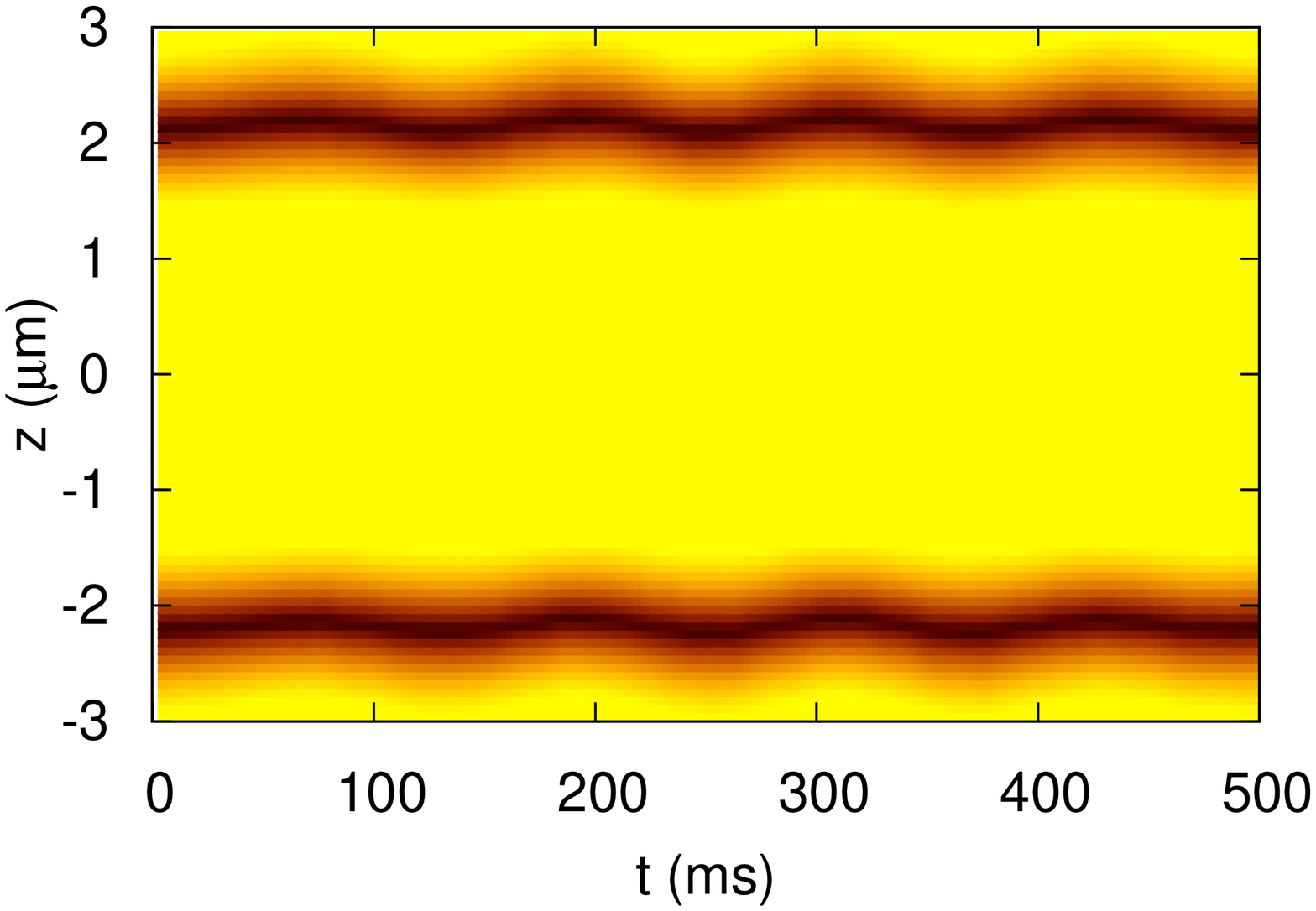}
\includegraphics[width=5.5cm,angle=270]{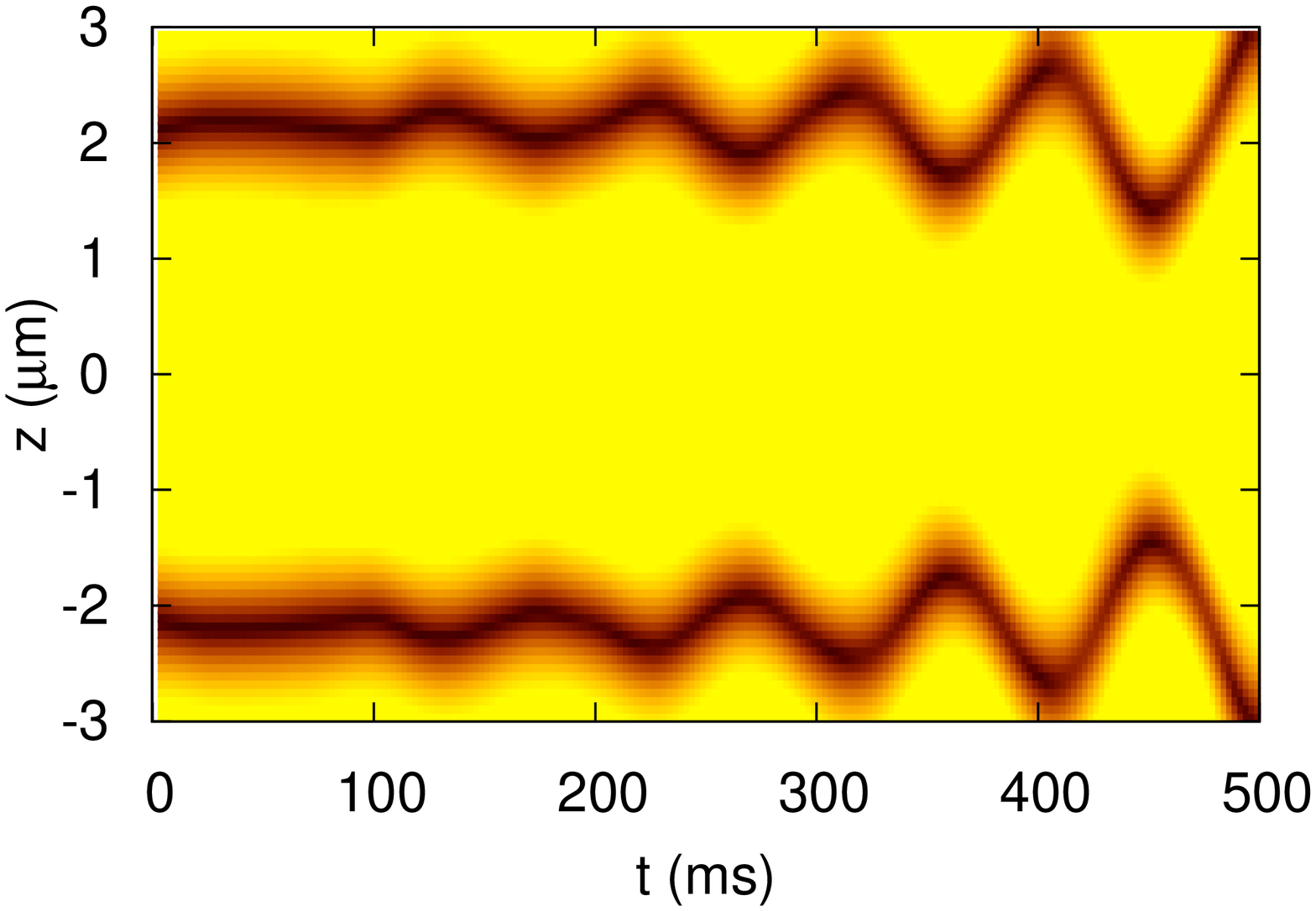}
\caption{(Color online) Same as in Fig.~\ref{Fig: 8}, but for the symmetric 
two-soliton state with in-phase (left) anomalous
mode and out-of-phase (right) imaginary
mode perturbations, for 
$\mu=3.1 \times 10^{-12}$ eV and $V_0=4.96 \times 10^{-13}$ eV.
The former anomalous eigenmode is stable, while the latter
is unstable (as concluded also from the BdG analysis).}
\label{Fig: 12}
\end{figure}

Fig.~\ref{Fig: 12} shows the time evolution of the symmetric 
two-soliton state for $\mu=3.1 \times 10^{-12}$ eV and 
$V_0=4.96 \times 10^{-13}$~eV perturbed by the eigenvector corresponding to the anomalous mode associated with in-phase (left panel) 
and by the eigenvector corresponding to the imaginary mode out-of-phase (right panel) oscillations of the dark solitons. 
For these parameters, the BdG analysis predicts that the former mode is 
stable while the  
latter is unstable. This is confirmed by the direct simulations: 
it is clearly observed that the amplitude of the out-of-phase 
oscillation increases whereas the amplitude of the in-phase oscillation remains constant.

\begin{figure}[htbp]
\includegraphics[width=5.5cm,angle=270]{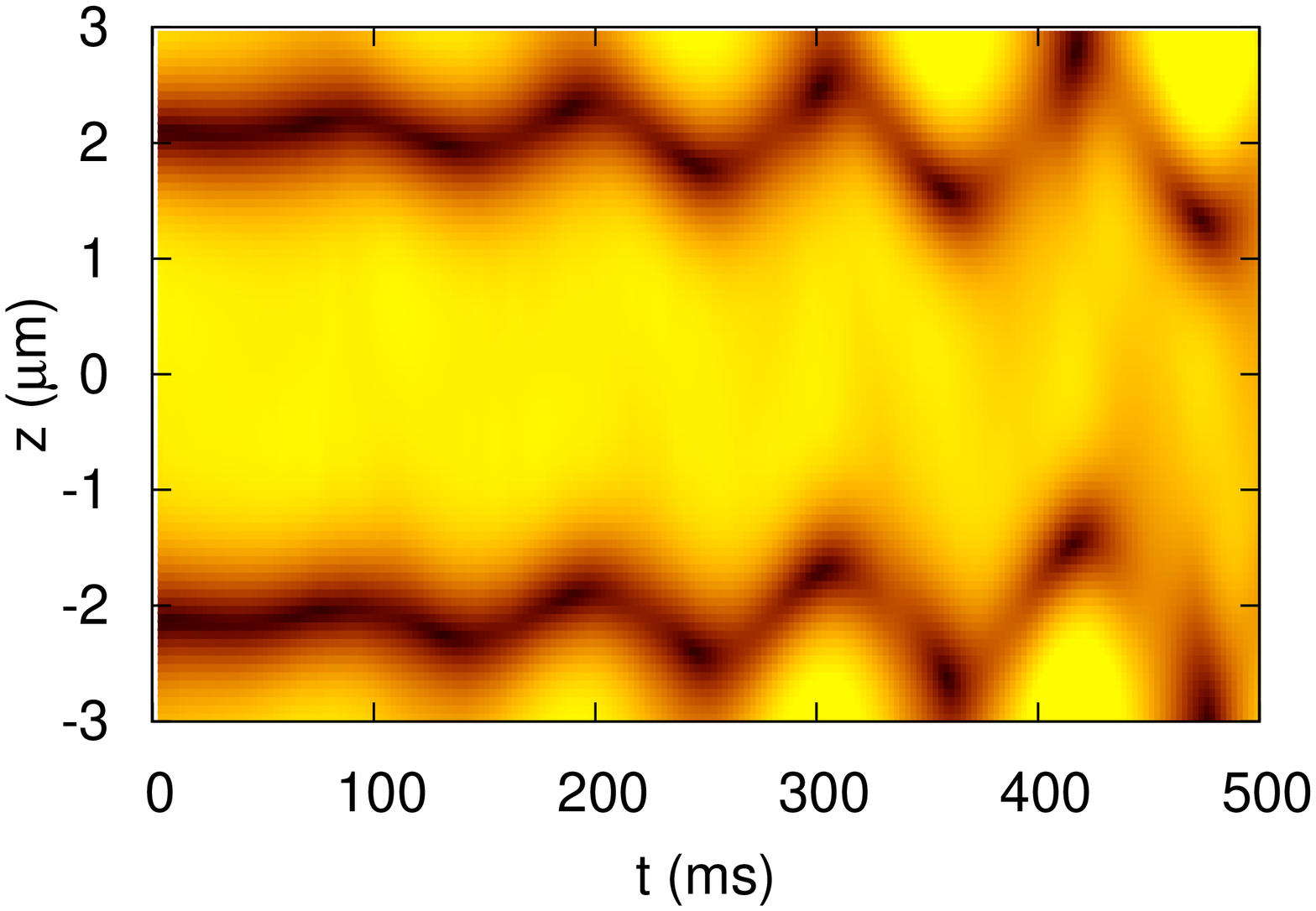}
\includegraphics[width=5.5cm,angle=270]{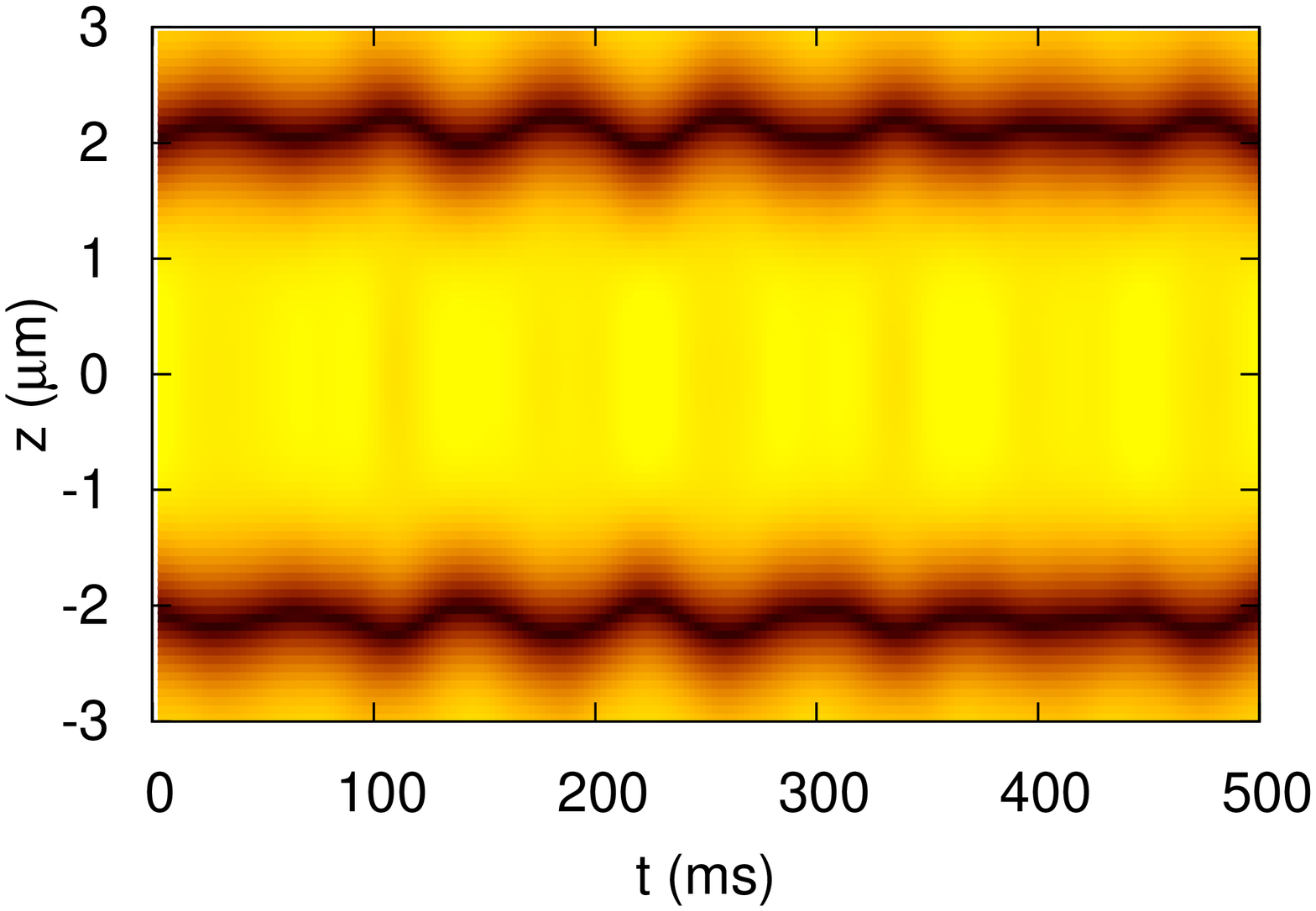}
\caption{(Color online) Same as in Fig.~\ref{Fig: 12}, but for the symmetric 
two-soliton state, for 
$\mu=2.8 \times 10^{-12}$ eV and $V_0=4.96 \times 10^{-13}$ eV.
Notice the reversal of stability of the two anomalous modes with respect
to Fig.~\ref{Fig: 12}.}
\label{Fig: 13}
\end{figure}

Finally, in Fig. \ref{Fig: 13} we show the time evolution of the symmetric two soliton state for 
$\mu=2.8 \times 10^{-12}$ eV and $V_0=4.96 \times 10^{-13}$ perturbed by the eigenvector corresponding to the anomalous mode 
with in-phase (left panel) and by the eigenvector corresponding to the imaginary mode out-of-phase (right panel) oscillations. 
For these parameters, the simulations show that 
the out-of-phase oscillation of the solitons is stable and the in-phase oscillation unstable, in agreement with the predictions of 
the BdG analysis.

\section{Conclusions}

In this work, we have performed a detailed study of the statics and 
dynamics of matter-wave dark solitons in a cigar-shaped 
Bose-Einstein condensate (BEC) confined in a double-well potential. 
The latter, was assumed to be formed as a combination of a usual 
harmonic trap and a periodic (optical lattice) potential. For our analysis, we 
have adopted and used an effectively 1D mean-field model, 
namely a Gross-Pitaevskii (GP)-like equation with a non-cubic nonlinearity, which has been used successfully in other works to describe dark solitons 
in BECs in the dimensionality crossover regime between 1D and 3D.

As a first step in our analysis, we studied the existence and stability of both the one- and multiple-dark soliton states. 
In particular, starting from the respective linear problem, we have used the 
continuation with respect to the chemical potential (number of atoms) to reveal 
the different branches of purely nonlinear solutions, including the ground 
state, as well as excited states, in the form of single or multiple dark 
solitons. 
We have shown that the presence of the optical lattice is crucial 
for the existence of nonlinear states that do not have a linear counterpart: 
these are actually asymmetric dark soliton states, with solitons located in one well (rather than in both wells) of the double-well potential, and do not 
exist in the non-interacting --- linear --- limit. We have systematically studied the bifurcations of the various branches of solutions and 
showed how each particular state emerges or disappears for certain chemical potential thresholds; the latter were determined analytically, 
in some cases, by using a Galerkin-type approach, in very good agreement with the numerical results. For each branch, we have also studied the 
stability of the pertinent solutions via a Bogoliubov-de Gennes (BdG) analysis. This way, we have discussed the role of the anomalous 
(negative energy) modes in the excitation spectra and revealed two types of 
instabilities of dark solitons, corresponding to either purely imaginary 
or genuinely complex eigenfrequencies; the latter were found to occur due to 
the collision of an anomalous mode with a positive energy mode and 
are associated with the so-called oscillatory instabilities of dark solitons.

We have also adopted a simple analytical model to study small-amplitude oscillations of the one- and multiple-dark-soliton states. Particularly, 
equations of motion for the single and multiple solitons were presented in the weakly-interacting limit (where the model becomes the usual 
cubic 1D GP equation), and modified them to take into regard the dimensionality of the system (i.e., the effect of transverse directions). The analysis 
of these equations of motion led to the determination of the characteristic soliton frequencies, which were then compared to the eigenfrequencies 
found in the framework of the BdG analysis; the agreement between the two 
was generally found to be very good. Perhaps even more importantly, the 
simple physical (i.e., particle) picture we adopted allowed us to reveal the 
role of the optical lattice strength (i.e., the height of the barrier in the 
double-well setting) 
in the stability of the dark solitons, as well as its role on the coupling 
between solitons. Particularly, it was found that sufficiently strong 
barriers lead to instability of symmetric and asymmetric soliton states, 
while it results in an effective decoupling of multi-solitons: the coupling 
between neighboring solitons as described by an effective repulsive 
potential, becomes negligible for sufficiently strong barriers. 
We should also note that generally the states possessing a dark soliton
at the center of the trap were found to be most strongly unstable, 
due to the emergence of a progressively larger (the larger the nonlinearity)
imaginary eigenfrequency.

Finally, we have performed systematic numerical simulations, based on direct 
numerical integration of the quasi-one-dimensional model, to investigate the 
dynamics of dark solitons in the double-well setup. We studied the manifestation of instabilities, when present, and found in all cases, 
a very good agreement between the predictions based on the 
BdG analysis and the numerical results. Instabilities have been observed
to manifest themselves through the exponential divergence of the soliton
from its initial location (when associated with an imaginary pair of
eigenfrequencies)
or the oscillatory growth (when associated with a complex quartet of
eigenfrequencies). 

An interesting direction for a future study would be a similar analysis in a 
higher-dimensional setting: this would refer to vortices confined 
in double-well potentials (or more complex ``energy surfaces'' such
as quadruple-well potentials \cite{chenyu2d}) 
in higher-dimensional BECs. Such studies are 
presently in progress.

\section*{Acknowledgments.} 

The work of D.J.F. was partially supported by the Special Account for Research Grants of the University of Athens. P.G.K. gratefully acknowledges
support from NSF-DMS-0349023 (CAREER) and NSF-DMS-0806762, as well as from 
the Alexander von Humboldt Foundation.


\end{document}